\documentclass[aps,prl,twocolumn,superscriptaddress]{revtex4-1}
\usepackage{amssymb, amsmath}
\usepackage{graphicx,onlyamsmath}
\usepackage[normalem]{ulem}  % для зачекивания текста

\usepackage{natbib}
\usepackage{xcolor}

\begin{document}

\title{Many-particle effects in optical transitions from zero-mode Landau levels in~HgTe~quantum wells}

\author{S.~S.~Krishtopenko}
%\thanks{These authors contributed equally to this work.}
\affiliation{Laboratoire Charles Coulomb, Universit\'{e} de Montpellier, Centre National de la Recherche Scientifique, 34095 Montpellier, France.}

\author{A.~M.~Kadykov}
%\thanks{These authors contributed equally to this work.}
\affiliation{Laboratoire Charles Coulomb, Universit\'{e} de Montpellier, Centre National de la Recherche Scientifique, 34095 Montpellier, France.}

\author{S.~Gebert}
%\thanks{These authors contributed equally to this work.}
\affiliation{Institut d'Electronique et des Syst\`emes, Universit\'{e} de Montpellier, Centre National de la Recherche Scientifique, 34000 Montpellier, France.}

\author{S.~Ruffenach}
\affiliation{Laboratoire Charles Coulomb, Universit\'{e} de Montpellier, Centre National de la Recherche Scientifique, 34095 Montpellier, France.}

\author{C.~Consejo}
\affiliation{Laboratoire Charles Coulomb, Universit\'{e} de Montpellier, Centre National de la Recherche Scientifique, 34095 Montpellier, France.}

\author{J.~Torres}
\affiliation{Institut d'Electronique et des Syst\`emes, Universit\'{e} de Montpellier, Centre National de la Recherche Scientifique, 34000 Montpellier, France.}

\author{C.~Avogadri}
\affiliation{Laboratoire Charles Coulomb, Universit\'{e} de Montpellier, Centre National de la Recherche Scientifique, 34095 Montpellier, France.}

\author{B.~Jouault}
\affiliation{Laboratoire Charles Coulomb, Universit\'{e} de Montpellier, Centre National de la Recherche Scientifique, 34095 Montpellier, France.}

\author{W.~Knap}
\affiliation{Laboratoire Charles Coulomb, Universit\'{e} de Montpellier, Centre National de la Recherche Scientifique, 34095 Montpellier, France.}

\author{N.~N.~Mikhailov}
\affiliation{Institute of Semiconductor Physics, Siberian Branch, Russian Academy of Sciences, pr. Akademika Lavrent'eva 13, Novosibirsk, 630090 Russia}
\affiliation{Novosibirsk State University, Pirogova st. 2, 630090 Novosibirsk, Russia.}

\author{S.~A.~Dvoretskii}
\affiliation{Institute of Semiconductor Physics, Siberian Branch, Russian Academy of Sciences, pr. Akademika Lavrent'eva 13, Novosibirsk, 630090 Russia}
\affiliation{Novosibirsk State University, Pirogova st. 2, 630090 Novosibirsk, Russia.}

\author{F.~Teppe}
\email[]{frederic.teppe@umontpellier.fr}
\affiliation{Laboratoire Charles Coulomb, Universit\'{e} de Montpellier, Centre National de la Recherche Scientifique, 34095 Montpellier, France.}
\date{\today}% It is always \today, today,

\begin{abstract}
We report on the far-infrared magnetospectroscopy of HgTe quantum wells with inverted band ordering at different electron concentrations. We particularly focus on optical transitions from zero-mode Landau levels, which split from the edges of electron-like and hole-like bands. We observe a pronounced dependence of the transition energies on the electron concentration varied by persistent photoconductivity effect. This is striking evidence that in addition to the already well-documented crystalline and interface asymmetries, electron-electron interactions also have a significant impact on the usual behavior of the optical transitions from zero mode Landau levels.
\end{abstract}

%Use showkeys class option if keyword                            %display desired
\maketitle

%\section{\label{sec:Int}Introduction}
HgTe/CdTe quantum wells (QWs) were the first two-dimensional (2D) systems in which it has been shown that the band ordering depended strongly on the QW width $d$~\cite{A1}. If $d$ is smaller than a critical value $d_c$, the first electron-like ($E1$) subband in the QW lies above the first hole-like ($H1$) subband, and the QW has a trivial band ordering~\cite{A2}. In wide QWs, when $d>d_c$, \emph{E}1 falls below \emph{H}1 and the band ordering becomes inverted, giving rise to the 2D topological insulator state~\cite{A2,A3}. At critical QW width, $d=d_c$, HgTe QWs host a gapless state with massless Dirac fermions~\cite{A4,A5,A6,A7}. The band ordering in HgTe QWs can also be changed by hydrostatic pressure~\cite{A8}, temperature~\cite{A9,A10} or strain~\cite{A11,A12}.

\begin{figure}
\includegraphics [width=1.0\columnwidth, keepaspectratio] {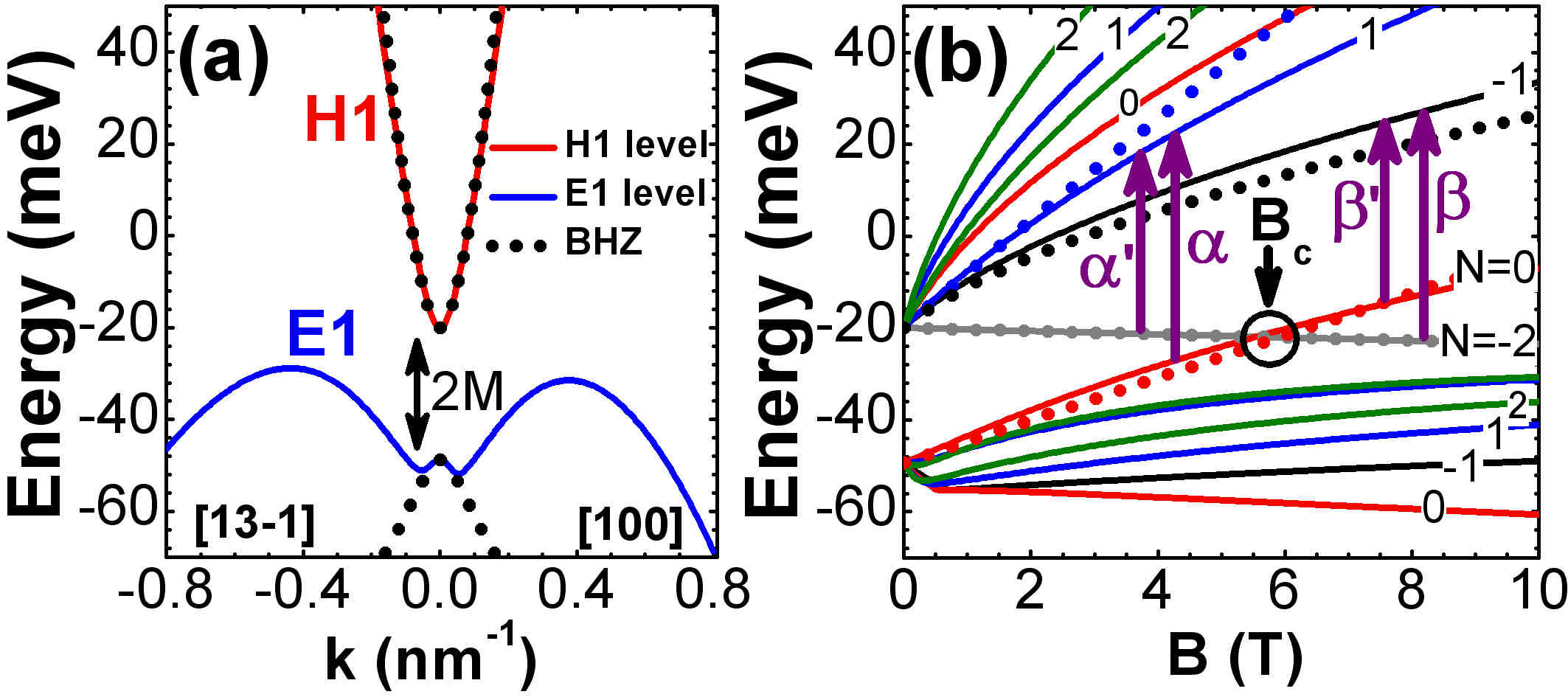} % Here is how to import EPS art
\caption{\label{Fig:1} Band structure (a) and Landau levels (b) of rectangular 8-nm-wide HgTe/Cd$_{0.7}$Hg$_{0.3}$Te QWs at $T=2$~K. (a) The blue and red curves represent band dispersion of $E1$ and $H1$ subbands, calculated within the 8-band \textbf{k$\cdot$p} Hamiltonian~\cite{A8}. The dotted curves show the dispersion within the BHZ model~\cite{A2}. (b) The numbers over the curves show the LL indices within the 8-band \textbf{k$\cdot$p} Hamiltonian~\cite{A8}. The arrows represent LL transitions observed in the vicinity of $B_c$~\cite{A7,A19,A20,A21}. The dotted colored curves show the corresponding LLs calculated within the BHZ model by using parameters provided in the Supplemental Materials~\cite{SM}.}
\end{figure}

The most efficient way to discriminate trivial and inverted band ordering in HgTe QWs is to probe the evolution of a particular pair of Landau levels (LLs) under applied magnetic field~\cite{A3}. These so-called \emph{zero-mode} LLs split from the edges of $E1$ and $H1$ subbands and have pure electron-like and hole-like character, respectively. The energy of the electron-like zero-mode LL increases systematically with magnetic field, while the energy of that of the \emph{H}1 subband decreases as the magnetic field increases. In inverted HgTe QWs, the zero-mode LLs therefore cross at a critical magnetic field $B_c$ (Fig.~\ref{Fig:1}), above which the inverted band ordering transforms into the trivial one~\cite{A3}.

The presence of bulk inversion asymmetry (BIA)~\cite{A13a,A13b} in the unit cell of zinc blende materials, as well as the interface inversion asymmetry (IIA) at the HgTe/CdHgTe heterojunction~\cite{A14} induce the anticrossing of zero-mode LLs in the vicinity of $B_c$. It appears that the value of this anticrossing gap $\Delta$ depends considerably on the experimental technique used to measure it. Particularly, the measurements of magnetotransport~\cite{A4,A5,A15,A16} and photoconductivity~\cite{A17,A18} performed with  gated Hall bars show that the anticrossing gap is negligibly small. On the contrary, the far-infrared magnetospectroscopy reveals a fine structure of the optical transitions from zero-mode LLs~\cite{A19,A20,A21,A22}. The analysis of this fine structure in the vicinity of $B_c$ within the Dirac-like model, including BIA and IIA, gives $\Delta\sim5$~meV for the joint effects. These contested experimental values of the anticrossing gap obtained in magnetotransport and magnetospectroscopy triggered a vivid discussion about the real strength of BIA and IIA in HgTe QWs~\cite{A5,A22,A23,A24}.

%The actual strength of BIA is also highly desirable to know for HgTe bulk crystals, as it results in a formation of 3D Weyl semimetal under compressive strain~\cite{A25}.

The key difference between magnetotransport and far-infrared magnetospectroscopy is that the latter induces inter-LL excitations, which may be considered as neutral collective modes~\cite{A26,A27} or \emph{magnetic excitons}~\cite{A27} composed of~a~bound state of a hole in a filled LL and an electron in an otherwise empty level. The long-wavelength limit of certain magnetic excitons~\cite{A26,A27}, such as magnetoplasmons, contributes into magnetooptical conductivity, defining the resonant energy and intensity of the magnetoabsorption lines~\cite{A29,A30}.

In 2D systems with parabolic band dispersion, all inter-LL transitions contributing into magnetoabsorption have the same cyclotron resonance (CR) energy, which is known to be unsensitive to electron-electron (\emph{e-e}) interaction~\cite{A28,A29}. Non-parabolic 2D systems have multiple LL transitions with different energies corresponding to multiple magnetoplamonic modes~\cite{A29,A30}. In such systems, the \emph{e-e} interaction mixes collective modes having close energies, already at zero wave-vector, making magnetooptical conductivity sensitive to many-particle effects~\cite{A33,A34}. So far, many-particle effects in magnetospectroscopy were observed in InAs QWs~\cite{A31,A32,A32b,A33,A34,A34b} and graphene~\cite{A35,A36,A37,A38}. As HgTe QWs also have strongly non-parabolic band structure~\cite{A2,A3}, many-particle effects should also contribute to their magnetooptical conductivity.

Here, we study the evolution of optical transitions from the zero-mode LLs in inverted HgTe QWs at different electron concentrations varied by the persistent photoconductivity effect~\cite{A39,A40}. By fitting the difference in the transition energies with an analytical expression including BIA and IIA, we extract the energy gap at the $\Gamma$ point of the Brillouin zone, the anticrossing gap $\Delta$ and the critical magnetic field $B_c$ from our experimental data. An unexpected strong dependence of the energy gap on the electron concentration clearly evidences that \emph{e-e} interaction affects the LL transitions beyond the single-particle picture.

%This paves the way towards clarifying the key reason for controversial conclusions on the strength of BIA and IIA in HgTe QWs obtained by magnetotransport and magnetospectroscopy.

Let us first consider the typical band structure and LLs of inverted HgTe QW in the absence of BIA and IIA (see Fig.~\ref{Fig:1}). The calculations were performed using the 8-band \textbf{k$\cdot$p} Hamiltonian~\cite{A8}. We also neglect the structure inversion asymmetry (SIA) assuming that the QW profile is symmetrical. To calculate the LLs, we apply the axial approximation by omitting the warping terms in the Hamiltonian~\cite{A8}. In this case, the electron-wave function for a given LL index $N>0$ generally has eight components, describing the contribution of the $\Gamma_6$, $\Gamma_7$, and $\Gamma_8$ bands into the LL. We note that a specific LL with $N=-2$ contains only a contribution of the heavy-hole band with a momentum projection $\pm3/2$~\cite{A8,A19,A20}. Details of the LL notation within the 8-band \textbf{k$\cdot$p} Hamiltonian are provided in Ref.~\cite{A8}.

The absence of BIA and IIA implies that the two zero-mode LLs, which can be recognized in the LLs with $N=-2$ and $N=0$ in Fig.~\ref{Fig:1}, simply cross each other at a critical magnetic field $B_c$~\cite{A8,A19,A20}. In this case, optically active inter-LL transitions follow conventional $\Delta N=\pm1$ selection rules imposed by angular momentum conservation law~\cite{A19}. Transitions from the zero-mode LLs, which follow these selection rules are marked in Fig.~\ref{Fig:1} as $\alpha$ and $\beta$ transitions, in accordance with the notation of Schultz \emph{et al.}~\cite{A42}. On the contrary, the $\alpha'$ and $\beta'$ transitions from the zero-mode LLs both correspond to "spin-flip" transitions~\cite{A3,A6}, which are forbidden in the single-particle picture if BIA and IIA are ignored.

The inter-LL transitions can be also analytically described  within the Dirac-like model proposed by Bernevig, Hughes and Zhang (BHZ)~\cite{A2,A3}. This BHZ model is directly derived from the 8-band \textbf{k$\cdot$p} Hamiltonian by applying a perturbation approach for the QW states in the vicinity of $\Gamma$ point~\cite{A2}. By using parameters provided in the Supplemental Materials~\cite{SM}, one can see that the BHZ model well describes the electronic states at small values of $\mathbf{k}$ (see Fig.~\ref{Fig:1}(a)). The colored dotted curves in Fig.~\ref{Fig:1}(b) show the energy of LLs involved in $\alpha$, $\alpha'$, $\beta$ and $\beta'$ transitions calculated within the BHZ model.

It is seen that only the energies of the zero-mode LLs are in good agreement with realistic numerical calculations. The difference in the energies calculated within both models does not exceed 10\% for the zero-mode LL from the $E1$ subband, while the final levels of $\alpha$, $\alpha'$, $\beta$ and $\beta'$ transitions show a significant deviation.
%Thus, one may seem that the BHZ model cannot be applied to describe evolution of these transitions in magnetic field.
However, by combining the energies of $\alpha$ with $\alpha'$ transition and $\beta$ with  $\beta'$ transition, we exclude the "wrong" LLs from consideration and apply the BHZ model to the energy difference:
\begin{equation}
\label{eq:2}
\Delta E=\dfrac{\hbar\omega_{\alpha'}-\hbar\omega_{\alpha}}{2}=\dfrac{\hbar\omega_{\beta'}-\hbar\omega_{\beta}}{2}=
\dfrac{\epsilon^{(+)}_0-\epsilon^{(-)}_0}{2},
\end{equation}
where $\epsilon^{(+)}_0$ and $\epsilon^{(-)}_0$ are the energies of the zero-mode LLs from the $E1$ and $H1$ subband, respectively. In the presence of SIA, BIA and IIA, the energies $\epsilon^{(\pm)}_0$ can be calculated analytically within the BHZ model~\cite{SM}:
\begin{equation}
\label{eq:3}
\Delta E=\sqrt{M^2\left(1-\dfrac{B}{B_c}\right)^2+\dfrac{\Delta^2}{4}},
\end{equation}
where $\Delta$ is the anticrossing gap at $B=B_c$ caused by both BIA and IIA, while $B_c$ and $M$ are the critical field and the mass parameter, respectively, both introduced in the absence of BIA and IIA~\cite{SM}. The parameter $M$ defines the gap between the $E1$ and $H1$ subbands at the $\Gamma$ point of the Brillouin zone (see Fig.~\ref{Fig:1}): it is positive for trivial band ordering and negative for inverted band structure. Thus, by fitting experimental values of the energy differences for both pairs of the transitions, one can directly extract the values of $\Delta$, $B_c$ and $M$ from magnetoabsorption. We note that, in contrast to the band structure shown in Fig.~\ref{Fig:1}, Eq.~(\ref{eq:3}) is also valid for asymmetrical QWs~\cite{SM}.

In this work, we have studied three different samples, each containing a 8-nm-wide HgTe QW embedded between Cd$_x$Hg$_{1-x}$Te barriers: $x=0.62$ for sample 091223, $x=0.41$ for sample 101221 and $x=0.77$ for sample 101109. The samples were grown by molecular beam epitaxy (MBE) on a semi-insulating (013) GaAs substrate with a relaxed CdTe buffer~\cite{A41}. The barriers were selectively doped with indium, resulting in a 2D electron concentration $n_S$ of a few $10^{11}$ cm$^{-2}$ at low temperatures. The magnetoabsorption spectra were measured in the Faraday configuration at 2~K by using a Fourier transform spectrometer coupled to a 16-T superconducting coil~\cite{A43}. All spectra were normalized by the sample transmission at zero magnetic field. In the measurements, the electron concentration was varied through the persistent photoconductivity effect~\cite{A39,A40} by changing the time of illumination with a blue light emitting diode (LED). We note that illumination of HgTe QWs with blue LED results in increasing of $n_S$~\cite{A40b}, in contrast to the case of InAs/GaSb QWs~\cite{A40}. The concentration values were determined via magnetotransport measurements in the van der Pauw geometry.

\begin{figure}
\includegraphics [width=1.0\columnwidth, keepaspectratio] {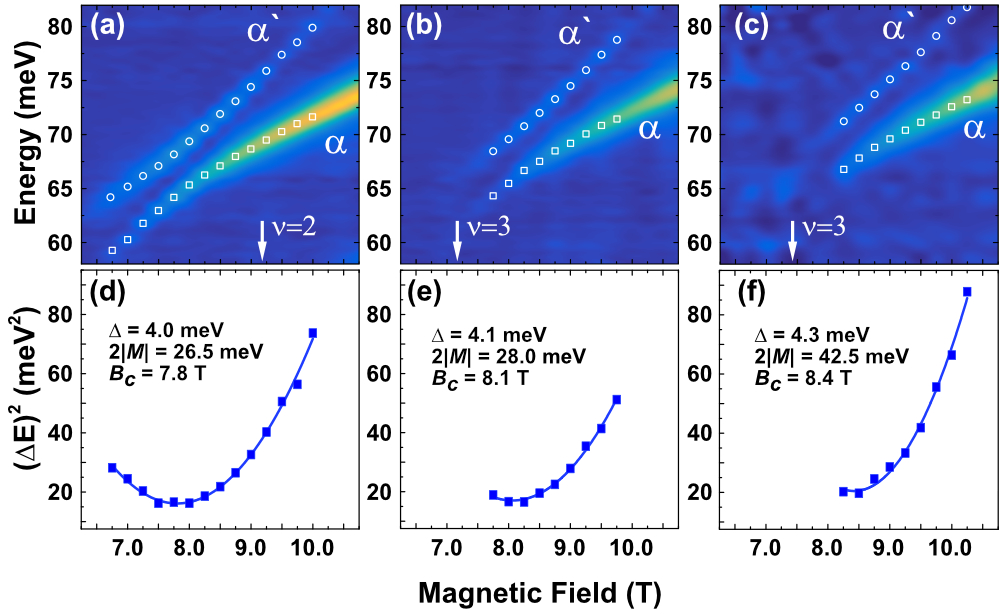} % Here is how to import EPS art
\caption{\label{Fig:2} (a)-(c) Color maps showing $\alpha$ and $\alpha'$ inter-LL transitions in the 8~nm HgTe/Cd$_{0.77}$Hg$_{0.27}$Te QW (sample 101109) as a function of magnetic field, measured at different electron concentration $n_S$: (a)~$4.5\cdot10^{11}$~cm$^{-2}$, (b)~$5.3\cdot10^{11}$~cm$^{-2}$, (c)~$5.7\cdot10^{11}$~cm$^{-2}$. The symbols represent position of the magnetoabsorption lines, whose energies are used in the evaluation of $\Delta E$. (d)-(f) Square of the energy difference for $\alpha$ and $\alpha'$ transitions at the same concentrations as in the respective top panels. The solid curves are the fitting to Eq.~(\ref{eq:3}). The arrows indicate magnetic fields corresponding to the integer filling factor $\nu$.}
\end{figure}

The magnetoabsorption spectra for the samples 101109 and 091223 are shown in Figs~\ref{Fig:2} and \ref{Fig:3}, respectively. The spectra of the sample 101221 are provided in the Supplemental Materials~\cite{SM}. As we are interested in the fine structure of the $\alpha$ and $\beta$ transitions in the vicinity of $B_c$, we only supply the high-energy parts of the spectra, above the reststrahlen band of the GaAs substrate (typically in the 30--40~meV range)~\cite{A19,A20}. The low-energy parts feature several CR-like transitions from higher LLs in the conduction band  and look qualitatively the same, as in previous works~\cite{A7,A19,A20,A21}.

The most intense line for the sample 101109 (see Fig.~\ref{Fig:2}) is identified as the $\alpha$ transition accompanied by the weaker line of the $\alpha'$ transition. As seen from Fig.~\ref{Fig:1}, the $\alpha$ and $\alpha'$ transitions are both present in the spectra if the filling factor of LLs $\nu$ in the conduction band is less than three. Indeed, $\nu<3$ for the $n_S$ values of $4.5\cdot10^{11}$~cm$^{-2}$, $5.3\cdot10^{11}$~cm$^{-2}$, $5.7\cdot10^{11}$~cm$^{-2}$ is fulfilled for magnetic fields higher then 6.2~T, 7.3~T and 7.9~T, respectively. The absence of the $\beta$ and $\beta'$ transitions in the spectra of the sample 101109 is attributed to $\nu>2$ in the given magnetic field range.

\begin{figure}
\includegraphics [width=1.0\columnwidth, keepaspectratio] {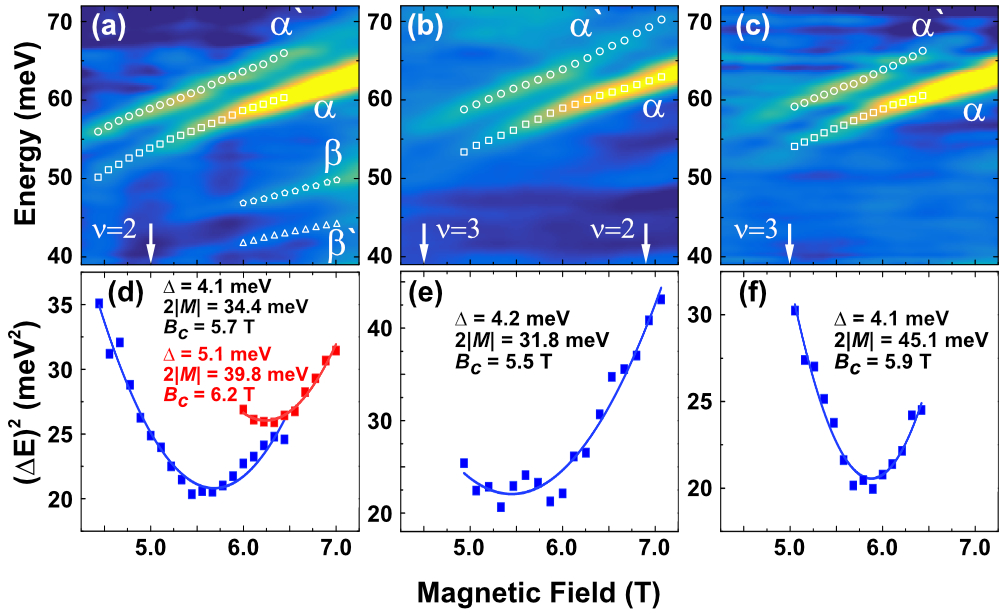} % Here is how to import EPS art
\caption{\label{Fig:3} (a)-(c) Color maps of the $\alpha$, $\alpha'$, $\beta$ and $\beta'$ inter-LL transitions as a function of magnetic field in the 8~nm HgTe/Cd$_{0.62}$Hg$_{0.38}$Te QW (sample 091223) at different electron concentration $n_S$: (a)~$2.4\cdot10^{11}$~cm$^{-2}$, (b)~$3.3\cdot10^{11}$~cm$^{-2}$, (c)~$3.6\cdot10^{11}$~cm$^{-2}$. The symbols correspond to the position of the magnetoabsorption lines, whose energies are used in the evaluation of $\Delta E$. (d)-(f) Square of the energy differences for both pairs of the transitions at the same concentrations as in the respective top panels. The solid curves are the fitting to Eq.~(\ref{eq:3}). The red and blue colors correspond to the $\alpha$-$\alpha'$ and $\beta$-$\beta'$ anticrossing, respectively. The arrows indicate magnetic fields corresponding to the integer filling factor $\nu$.}
\end{figure}

As seen from Fig.~\ref{Fig:3}(a), $n_S=2.4\cdot10^{11}$~cm$^{-2}$ allows for observation of all four $\alpha$, $\alpha'$, $\beta$ and $\beta'$ transitions in the sample 091223 since $\nu<2$ represents the fields higher than 5.0~T. Increasing of $n_S$ up to $3.3\cdot10^{11}$~cm$^{-2}$ yields to the vanishing of the $\beta$ and $\beta'$ transitions in the field range of $4.5$~T$<B<6.8$~T since it corresponds to $2<\nu<3$. We note that the  $\alpha'$ and $\beta'$ transitions are observed just in the vicinity of $B_c$, while above the field range shown in Fig.~\ref{Fig:3}, only the $\alpha$ and $\beta$ transitions are present.

\begin{figure}
\includegraphics [width=1.0\columnwidth, keepaspectratio] {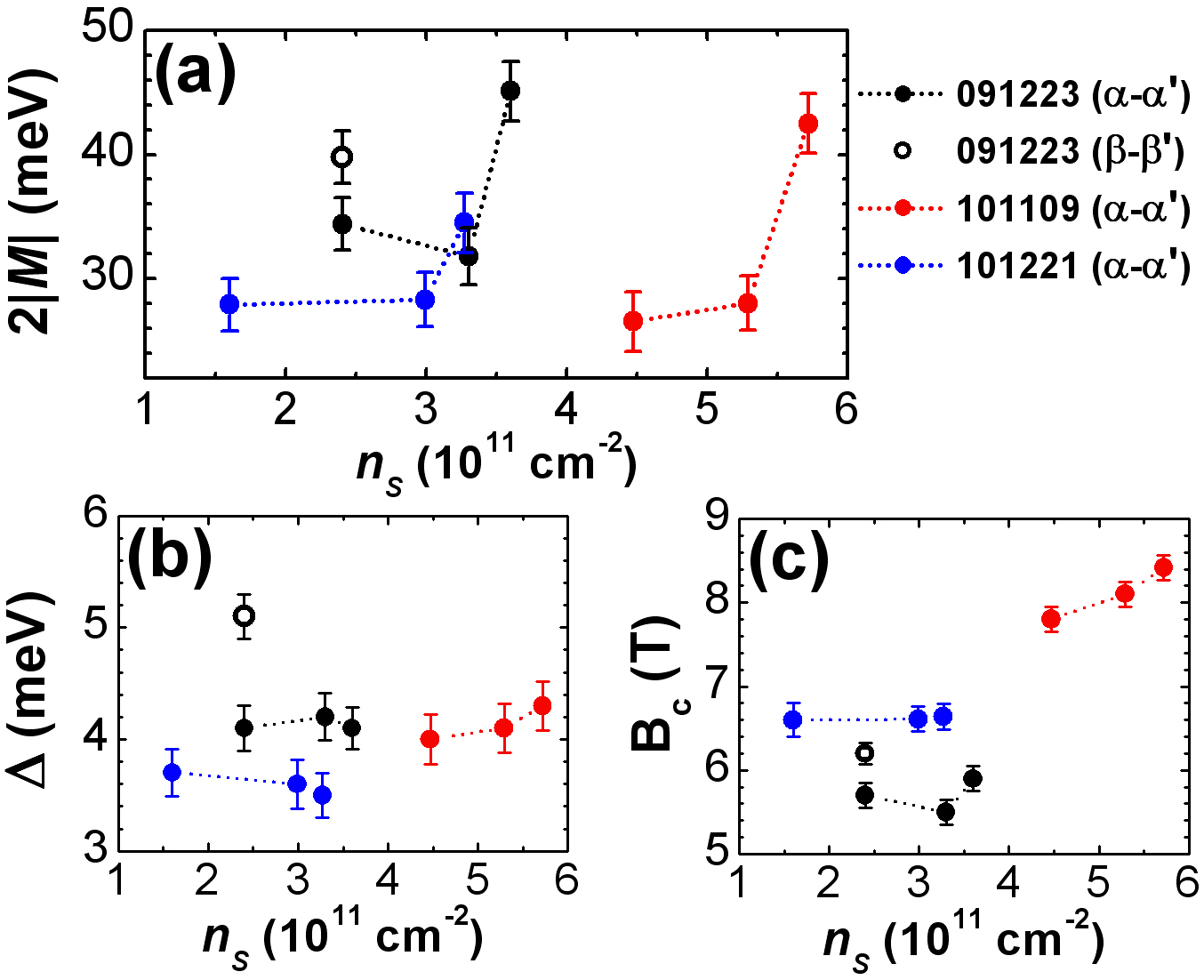} % Here is how to import EPS art
\caption{\label{Fig:4} Evolution of single-particle parameters with electron concentration extracted from magnetoabsorption of different samples: (a) $2|M|$ vs $n_S$, (b) $B_c$ vs $n_S$, (c) $\Delta$ vs $n_S$.}
\end{figure}

In order to analyze our magnetoabsorption data within the single-particle picture, we have fitted the difference in energies $\Delta E$ between $\alpha$ and $\alpha'$, and between $\beta$ and $\beta'$ transitions by Eq.~(\ref{eq:3}). As seen from Figs~\ref{Fig:2} and \ref{Fig:3}, the energy difference is formally well described by the BHZ model including the SIA, BIA and IIA effects. Figure~\ref{Fig:4} summarizes the values of $\Delta$, $B_c$ and $M$ as a function of $n_S$ for both pairs of the transitions. The error bar for the extracted values does not exceed 10\%.

In the following, we address some interesting features, which cannot be explained within the single-particle picture. First, the values of $\Delta$, $B_c$ and $M$ extracted from $\hbar\omega_{\alpha'}-\hbar\omega_{\alpha}$ and $\hbar\omega_{\beta'}-\hbar\omega_{\beta}$ differ significantly from each other. The difference between $\hbar\omega_{\alpha'}-\hbar\omega_{\alpha}$ and $\hbar\omega_{\beta'}-\hbar\omega_{\beta}$ is clearly seen in Fig.~\ref{Fig:3}(d). Note that in the single-particle picture, the energy differences $\hbar\omega_{\alpha'}-\hbar\omega_{\alpha}$ and $\hbar\omega_{\beta'}-\hbar\omega_{\beta}$ should be the same. This is a general property of the single-particle approach, which is still valid in the BHZ model~\cite{A21}.

Second, Fig.~\ref{Fig:4}(a) demonstrates a pronounced dependence of the energy gap $2M$ on the electron concentration. The changing range for $2M$ exceeds significantly the error bar and the deviation within 10\%, expected for zero-mode LLs in the BHZ model (see Fig.~\ref{Fig:1}). On the other hand, $\Delta$ and $B_c$ are either independent of $n_S$ or have a weak concentration dependence within the error bar. Although, the changing of $n_S$ affects the band structure via the changes of the QW profile, such effect is small in our samples. Particularly, the self-consistent calculations involving the Poisson equation and the 8-band \textbf{k$\cdot$p} Hamiltonian predict less than 5\% changing of $2M$ for pure asymmetrical QWs in the range of $n_S$ shown in Fig.~\ref{Fig:4}. Thus, the strong concentration dependence of $2|M|$ together with the different values of $\Delta$ and $B_c$ extracted from $\hbar\omega_{\alpha'}-\hbar\omega_{\alpha}$ and $\hbar\omega_{\beta'}-\hbar\omega_{\beta}$ cannot be interpreted within the single-particle picture.

Let us now discuss qualitatively possible mechanism beyond the single particle picture, which may result in the fine structure of $\alpha$ and $\beta$ transitions shown in Figs~\ref{Fig:2} and \ref{Fig:3}. As mentioned before, any inter-LL transition observed in magnetoabsorption can be considered as neutral magnetic exciton, which long-wavelength limit contributes to the magnetooptical conductivity~\cite{A26,A27,A29,A30}. In this sense, two LL transitions with close energies, such as $\alpha$ and $\alpha'$, correspond to two magnetic excitons with zero wave-vectors. In the absence of \emph{e-e} interaction, their energies are defined by the single-particle LLs and only the $\alpha$ exciton contributes in magnetoabsorption.

The many-particle interaction~\cite{A26,A27,A29,A30} gives rise to (1) an electron-hole interaction inside a given exciton, (its bound energy); (2) interaction between excited exciton and other non-excited electrons below the Fermi level; and (3) the exciton-exciton interaction caused by the interaction between the electrons and holes of the $\alpha$ and $\alpha'$ excitons. The first two interactions just change the energies of the $\alpha$ and $\alpha'$ excitons from its single-particle values. On the contrary, the interaction (3) induces hybridization between two magnetic excitons, which leads to non-zero $\alpha'$ contribution into magnetoabsorption and anticrossing between $\alpha$ and $\alpha'$ magneto-optical transitions.

This hybridization between the excitons is very efficient if the energies of $\alpha$ and $\alpha'$ transitions differ in less than the characteristic length of Coulomb interaction $E_c$ in our samples. The latter can be roughly evaluated as $E_c\sim e^2/(a_B\epsilon)$, where $e$ is an elementary charge, $a_B$ is the magnetic length given by $a_B^2=\hbar c/eH$ and $\epsilon=21$ is the static permittivity of HgTe. In the field range of $B=4-9$~T, $E_c$ changes from 5 to 8~meV, that is comparable with the experimental values of $\hbar\omega_{\alpha'}-\hbar\omega_{\alpha}$ and $\hbar\omega_{\beta'}-\hbar\omega_{\beta}$ in the vicinity of $B_c$. Therefore, it is relevant to account many-particle effects in the consideration of the fine structure of $\alpha$ and $\beta$ transitions. Thus, the strong dependence of the band-gap energy in Fig.~\ref{Fig:4} is due to the inapplicability in the vicinity of $B_c$ of the single particle model described by Eq.~(\ref{eq:3}).

Moreover, since the many-particle hybridization is sensitive to the electron concentration $n_S$ and LL filling factor $\nu$, it may indeed result in different fine structures for the $\alpha$ and $\beta$ transitions. Additionally, the proposed mechanism does not require the anticrossing of zero-mode LLs that is consistent with experimental evidences of the small values of BIA and IIA obtained by magnetotransport~\cite{A4,A5,A15,A16} and photoconductivity~\cite{A17,A18}.

Finally, we note that the previous magnetospectroscopy studies of HgTe QWs~\cite{A6,A7,A19,A20} have shown a good agreement between experimental values and single-electron calculations for all observed transitions in \emph{trivial} QWs~\cite{A6,A7,A20} and in \emph{inverted} QWs in the field range far from $B_c$~\cite{A7,A19,A20}. In these cases, the difference between the transition energies is greater than $E_c\sim e^2/(a_B\epsilon)$. The latter means that unlike in graphene, in which the LL transitions are affected by many-particle interaction in the whole range of magnetic fields~\cite{A35,A36,A37,A38}, unhybridized optical transitions in HgTe QWs can be treated far from the critical field $B_c$ within the single-electron picture.

In conclusion, we have studied inverted HgTe/CdHgTe QWs by far infrared magneto-spectroscopy, by varying the carrier density with a persistent photoconductivity effect. The single-electron analysis of several optical transitions from the zero-mode LLs near their crossing, highlights the contribution of many-particles phenomena, via an unexpectedly strong dependence of the band gap energy as a function of the electron concentration. This indicates that LL transitions from zero-mode LLs probed by far-infrared magnetospectroscopy should be considered in terms of magnetic excitons, as collective modes~\cite{A26,A27,A29,A30,A31,A33,A34}, hybridized by many-particle interaction.

%~~***~~\cite{SM6,SM7,SM14}

\begin{acknowledgments}
The authors gratefully thank M. Orlita (Laboratoire National des Champs Magnetiques Intenses, Grenoble) for the helpful discussions and critical comments. This work was supported by MIPS department of Montpellier University through the "Occitanie Terahertz Platform", by CNRS through IRP "TeraMIR", by the French Agence Nationale pour la Recherche (Colector project), and by the European Union through the Flag-Era JTC 2019 - DeMeGras project and the Marie-Curie grant agreement No~765426, from Horizon 2020 research and innovation programme.
\end{acknowledgments}

%\bibliography{Anticrossing2}
%

\newpage
\clearpage
\setcounter{equation}{0}
\setcounter{figure}{0}
\setcounter{table}{0}
\renewcommand{\thefigure}{S\arabic{figure}}

\onecolumngrid
\section*{Supplemental Materials}
\maketitle
\onecolumngrid

\subsection{A. Dirac-like 2D Hamiltonian}
The low-energy electronic states in HgTe quantum wells (QWs) in the vicinity of the $\Gamma$ point of the Brillouin zone can be described analytically within an effective four-band Dirac-like model. This model, first proposed by Bernevig, Hughes and Zhang~\cite{SM1} considers the lowest electron-like level \emph{E}1 and the top hole-like level \emph{H}1 that qualitatively explains a topological phase transition in 2D systems with spatial inversion symmetry~\cite{SM1}. In the basis $|E1,+\rangle$, $|H1,+\rangle$, $|E1,-\rangle$, $|H1,-\rangle$ and keeping terms only up to quadratic powers of momentum $\mathbf{k}$, the Bernevig-Hughes-Zhang (BHZ) Hamiltonian is written as
\begin{equation}\label{eq:SM1}
H_{\mathrm{BHZ}}(\mathbf{k}) = \begin{pmatrix}
   C+M-(\mathbb{D}+\mathbb{B})k^2 & Ak_{+}  & 0 & 0\\
   Ak_{-} & C-M-(\mathbb{D}-\mathbb{B})k^2  & 0 & 0\\
    0 & 0  & C+M-(\mathbb{D}+\mathbb{B})k^2 & -Ak_{-}\\
    0 & 0  & -Ak_{+} & C-M-(\mathbb{D}-\mathbb{B})k^2
    \end{pmatrix},
\end{equation}
where the upper 2$\times$2 block describes spin-up electrons in the \emph{E}1 and \emph{H}1 subbands, and the lower block corresponds to the spin-down states in those subbands~\cite{SM1}.
Here, $\mathbf{k}=(k_x,k_y)$ are the momentum components in the QW plane and $k_{\pm}=k_x\pm ik_y$. The energy gap between the bands is $2M$, $\mathbb{B}$ and $\mathbb{D}$ describes the curvature of the subbands, while $A$ incorporates inter-subband coupling to lowest order.
The negative values of $M$ correspond to the inverted band structure, while $M>0$ describes the trivial band ordering.

However, real HgTe QWs do not have inversion symmetry for several reasons, that leads to additional terms in the Hamiltonian. If the QW has a structure inversion asymmetry (SIA) in the growth direction, the SIA term reads~\cite{SM2}
\begin{equation}\label{eq:SM2}
H_{\mathrm{SIA}}(\mathbf{k}) = \begin{pmatrix}
   0 & 0  & i\alpha k_{-} & 0\\
   0 & 0  & 0 & 0\\
    -i\alpha k_{+} & 0  & 0 & 0\\
    0 & 0  & 0 & 0
    \end{pmatrix},
\end{equation}
which is simply the Rashba term linear in $k$; the heavy-hole $k$-cubic Rashba term is neglected. The bulk inversion asymmetry (BIA) arising due to the presence of two different atoms in the unit cell of the layer materials indices the following BIA terms~\cite{SM2,SM3}:
\begin{equation}\label{eq:SM3}
H_{\mathrm{BIA}}(\mathbf{k}) = \begin{pmatrix}
   0 & 0  & \Delta_{e}k_{+} & -\Delta_{0}\\
   0 & 0  & \Delta_{0} & \Delta_{h}k_{-}\\
    \Delta_{e}k_{-} & \Delta_{0}  & 0 & 0\\
    -\Delta_{0} & \Delta_{h}k_{+}  & 0 & 0
    \end{pmatrix}.
\end{equation}
Finally, the anisotropy of the QW interfaces induces an interface inversion asymmetry (IIA) leading to efficient interface coupling between the light-hole $|\Gamma_8,\pm1/2\rangle$ and heavy-hole $|\Gamma_8,\pm1/2\rangle$ states~\cite{SM4}. When projected onto the \emph{E}1 and \emph{H}1 subbands the IIA terms are
\begin{equation}\label{eq:SM4}
H_{\mathrm{IIA}} = \begin{pmatrix}
   0 & 0  & 0 & \gamma\\
   0 & 0  & -\gamma & 0\\
    0 & -\gamma  & 0 & 0\\
    \gamma & 0  & 0 & 0
    \end{pmatrix}.
\end{equation} The parameters $C$, $A$, $\mathbb{B}$, $\mathbb{D}$, $M$, $\alpha$, $\Delta_{e}$, $\Delta_{h}$, $\Delta_{0}$, $\gamma$ in Eqs.~(\ref{eq:SM1}), (\ref{eq:SM2}), (\ref{eq:SM3}), (\ref{eq:SM4}) depend on the QW geometry, QW width and the layer materials. Thus, the Hamiltonian for HgTe QWs naturally separates into three parts:
\begin{equation}\label{eq:SM5}
H_D(\mathbf{k})=H_{\mathrm{BHZ}}(\mathbf{k})+H_{\mathrm{SIA}}(\mathbf{k})+H_{\mathrm{BIA}}(\mathbf{k})+H_{\mathrm{IIA}}.
\end{equation}

\subsection{B. Zero-mode Landau levels in the presence of SIA, BIA and IIA}

To calculate Landau levels (LLs) in the presence of an external magnetic field $B$ oriented perpendicular to the QW plane, one should make the Peierls substitution, $\hbar\mathbf{k}\rightarrow\hbar\mathbf{k}-\frac{e}{c}\mathbf{A}$, where in the Landau gauge $\mathbf{A}=B(y,0)$. Then we can replace the momentum operators in Eqs.~(\ref{eq:SM1}), (\ref{eq:SM2}), (\ref{eq:SM3}), (\ref{eq:SM4}) with standard ladder operators $\hat{a}^{+}$ and $\hat{a}$:
\begin{equation}
\label{eq:SM5}
k_{+}\rightarrow\dfrac{\sqrt{2}}{a_B}\hat{a}^{+},~~~~
k_{-}\rightarrow\dfrac{\sqrt{2}}{a_B}\hat{a},~~~~
k^{2}\rightarrow\dfrac{2}{a_B^2}\left(\hat{a}^{+}\hat{a}+\dfrac{1}{2}\right),
\end{equation}
where $a_B$ is the magnetic length given by $a_B^2=\hbar c/eB$. To take into account the Zeeman splitting, we also add an additional term to the Hamiltonian $H_D(\mathbf{k})$:
\begin{equation}
\label{eq:SM6}
H_{Z}=\dfrac{\mu_BB}{2}\begin{pmatrix}
g_e & 0 & 0 & 0 \\
0 & g_h & 0 & 0 \\
0 & 0 & -g_e & 0 \\
0 & 0 & 0 & -g_h\end{pmatrix},
\end{equation}
where $\mu_B$ is the Bohr magneton, $g_e$ and $g_h$ are the effective (out-of-plane) g-factors of the \emph{E}1 and \emph{H}1 subbands, respectively.

Let us first consider, LLs in the absence of SIA, BIA and IIA. In this case, $H_{\mathrm{BHZ}}(\mathbf{k})+H_{Z}$ has a block-diagonal form, and the eigenvalue problem can be solved analytically~\cite{SM5}. The upper block gives the following LL energies $E^{(+)}_{n,s}$
\begin{equation*}
E^{(+)}_{n,s}=C-\dfrac{2\mathbb{D}n+\mathbb{B}}{a_B^2}+\dfrac{g_e+g_h}{4}\mu_BB
+s\sqrt{\dfrac{2nA^2}{a_B^2}+\left(M-\dfrac{2\mathbb{B}n+\mathbb{D}}{a_B^2}+\dfrac{g_e-g_h}{4}\mu_BB\right)^2}, \text{~~~~~~~for $n\geq1$}
\end{equation*}
\begin{equation}
\label{eq:SM7}
E^{(+)}_0=C+M-\dfrac{\mathbb{D}+\mathbb{B}}{a_B^2}+\dfrac{g_e}{2}\mu_BB, \text{~~~~~~~for $n=0$},
\end{equation}
where $n$ is LL index and $s=\pm$  represents LLs in conduction and valence subband. Explicitly, the two-component eigenfunctions for the upper block are given by
\begin{equation}
\label{eq:SM8}
\Psi^{(+)}_{n,s}=\begin{pmatrix}
\cos \theta^{(+)}_{n,s} |n\rangle \\
\sin \theta^{(+)}_{n,s} |n-1\rangle   \\
0  \\
0 \end{pmatrix},
\end{equation}
where $|n\rangle$ is the normalized harmonic oscillator function~\cite{SM6} and $\theta^{(+)}_{n,s}$ is defined as
\begin{equation*}
\tan \theta^{(+)}_{n,s}=\dfrac{\frac{\sqrt{2}\hbar}{a_B}A\sqrt{n}}
{E^{(+)}_{n,s}-C+M+\frac{2\hbar^2}{a_B^2}(\mathbb{D}-\mathbb{B})\left(n-\frac{1}{2}\right)-\frac{g_h\mu_BB}{2}}.
\end{equation*}
Note that for $n=0$, $\theta^{(+)}_{n=0,s}=0$.

For the lower block of $H(\mathbf{k})+H_{Z}$, the LL energies $E^{(-)}_n$ are written as
\begin{equation*}
E^{(-)}_n=C-\dfrac{2\mathbb{D}n-\mathbb{B}}{a_B^2}-\dfrac{g_e+g_h}{4}\mu_BB
+s\sqrt{\dfrac{2nA^2}{a_B^2}+\left(M-\dfrac{2\mathbb{B}n-\mathbb{D}}{a_B^2}-\dfrac{g_e-g_h}{4}\mu_BB\right)^2}, \text{~~~~~~~for $n\geq1$}
\end{equation*}
\begin{equation}
\label{eq:SM9}
E^{(-)}_0=C-M-\dfrac{\mathbb{D}-\mathbb{B}}{a_B^2}-\dfrac{g_h}{2}\mu_BB, \text{~~~~~~~for $n=0$}.
\end{equation}
The two-component eigenfunctions for the upper block have the form
\begin{equation}
\label{eq:SM10}
\Psi^{(-)}_{n,s}=\begin{pmatrix}
0 \\
0 \\
-\sin\theta^{(-)}_{n,s}|n-1\rangle  \\
\cos\theta^{(-)}_{n,s}|n\rangle  \end{pmatrix},
\end{equation}
where $\theta^{(-)}_{n,s}$ is defined as
\begin{equation*}
\tan \theta^{(-)}_{n,s}=\dfrac{\frac{\sqrt{2}\hbar}{a_B}A\sqrt{n}}
{E^{(-)}_{n,s}-C-M+\frac{2\hbar^2}{a_B^2}(\mathbb{D}+\mathbb{B})\left(n-\frac{1}{2}\right)+\frac{g_e\mu_BB}{2}}.
\end{equation*}

The LLs with energies $E^{(+)}_0$ and $E^{(-)}_0$ are called the \emph{zero-mode} LLs~\cite{SM5}. They split from the edge of \emph{E}1 and \emph{H}1 subbands and tend toward conduction and valence band as a function of $B$, respectively.  The critical magnetic field $B_c$, where those two LLs cross, can be calculated from the condition $E^{(+)}_0= E^{(-)}_0$, which yields
\begin{equation}
\label{eq:SM11}
B_c=\dfrac{M}{e\mathbb{B}/(c\hbar)-(g_e+g_h)\mu_B/4}.
\end{equation}

By applying the procedure described in Ref.~\cite{SM7} to the 8-band \textbf{k$\cdot$p} Hamiltonian, we have found that $A=348.1$~meV$\cdot$nm, $\mathbb{B}=-1051.2$~meV$\cdot$nm$^2$, $C=-34.6$~meV, $\mathbb{D}=-868.1$~meV$\cdot$nm$^2$, $g_e=58.50$, $g_h=2.41$ and $M=-14.6$~meV for 8-nm-wide HgTe/Cd$_{0.7}$Hg$_{0.3}$Te QW grown on CdTe buffer at 2~K (see Fig.~1 in the main text).
The presence of SIA, BIA and IIA breaks a block-diagonal form of the Hamiltonian due to the non-diagonal contribution
\begin{equation}\label{eq:SM12}
H_{\mathrm{SIA}}(\mathbf{k})+H_{\mathrm{BIA}}(\mathbf{k})+H_{\mathrm{IIA}} = \begin{pmatrix}
   0 & 0  & \Delta_{e}\frac{\sqrt{2}\hbar}{a_B}\hat{a}^{+}+i\alpha\frac{\sqrt{2}\hbar}{a_B}\hat{a} & \gamma-\Delta_{0}\\
   0 & 0  & -\gamma+\Delta_{0} & \Delta_{h}\frac{\sqrt{2}\hbar}{a_B}\hat{a}\\
    \Delta_{e}\frac{\sqrt{2}\hbar}{a_B}\hat{a}-i\alpha\frac{\sqrt{2}\hbar}{a_B}\hat{a}^{+} & -\gamma+\Delta_{0}  & 0 & 0\\
    \gamma-\Delta_{0} & \Delta_{h}\frac{\sqrt{2}\hbar}{a_B}\hat{a}^{+}  & 0 & 0
    \end{pmatrix}.
\end{equation}
However, the energies of zero-mode LLs can be calculated analytically even in this case. Indeed, with the wave function
\begin{equation}
\label{eq:SM11}
\Psi^{(\pm)}_{0}=\begin{pmatrix}
a|0\rangle \\
0 \\
0  \\
b|0\rangle  \end{pmatrix},
\end{equation}
where $a$ and $b$ are the constants, the problem of calculating the energy levels, $(H_D(\mathbf{k})+H_{Z})\Psi^{(\pm)}_{0}=\epsilon^{(\pm)}_{0}\Psi^{(\pm)}_{0}$ gives
\begin{equation}
\label{eq:SM12}
\epsilon^{(\pm)}_0=C-\dfrac{\mathbb{D}}{a_B^2}+\dfrac{g_e-g_h}{4}\mu_BB\pm
\sqrt{\left(M-\dfrac{\mathbb{B}}{a_B^2}+\dfrac{g_e+g_h}{4}\mu_BB\right)^2+\left(\Delta_{0}-\gamma\right)^2},
\end{equation}
which is independent of $\alpha$, $\Delta_{e}$ and $\Delta_{h}$.
By combining Eq.~(\ref{eq:SM12}) with the expression for $B_c$ given by Eq.~(\ref{eq:SM11}), it is worth to present $\epsilon^{(\pm)}_0$ in the form:
\begin{equation}
\label{eq:SM13}
\epsilon^{(\pm)}_0=C-\dfrac{\mathbb{D}}{a_B^2}+\dfrac{g_e-g_h}{4}\mu_BB\pm
\sqrt{M^2\left(1-\dfrac{B}{B_c}\right)^2+\left(\Delta_{0}-\gamma\right)^2}.
\end{equation}
One can see that both BIA and IIA induces the anticrossing of zero-mode LLs with the anticrossing gap $\Delta=2\left|\Delta_{0}-\gamma\right|$ at $B=B_c$. Thus, Eq.~(\ref{eq:SM13}) transforms into the given expression for the energy difference:
\begin{equation}
\label{eq:SM14}
\Delta E=\dfrac{\epsilon^{(+)}_0-\epsilon^{(-)}_0}{2}=\sqrt{M^2\left(1-\dfrac{B}{B_c}\right)^2+\dfrac{\Delta^2}{4}}.
\end{equation}
We note that Eq.~(\ref{eq:SM14}) has been derived by taking into account the SIA effect, therefore, it holds for the asymmetrical QW as well.

\subsection{C. Changing of the QW profile with electron concentration}
The "optical gating", also known as persistent photoconductivity effect~\cite{SM11,SM12,SM13}, is caused by the charge transfer between the QW and the cap layer, in which the illumination induces either an ionization of the deep donors, or inter-band excitations~\cite{SM13}. However, electric field of spatially separated donors in the cap layer and 2D electrons in HgTe QW distorts the QW profile. Therefore, by varying the electron concentration $n_S$, one also changes the "built-in" electric field and thus asymmetry of the QW. The latter affects the band structure parameters, such as the band-gap at the $\Gamma$ point of the Brillouin zone $2M$ and critical magnetic field $B_c$.

In order to evaluate effect of QW distortion on $M$ and $B_c$, we have performed self-consistent band structure and LLs calculations, based on the joint solution of Poisson and Schr\"{o}dinger equations with the 8-band \textbf{k$\cdot$p} Hamiltonian. Such method allows to take into account asymmetry of the HgTe QW in the right way, which gives a good agreement with magnetotransport results obtained for the samples with the gated Hall bars~\cite{SM14}.

\begin{figure}
\includegraphics [width=0.95\columnwidth, keepaspectratio] {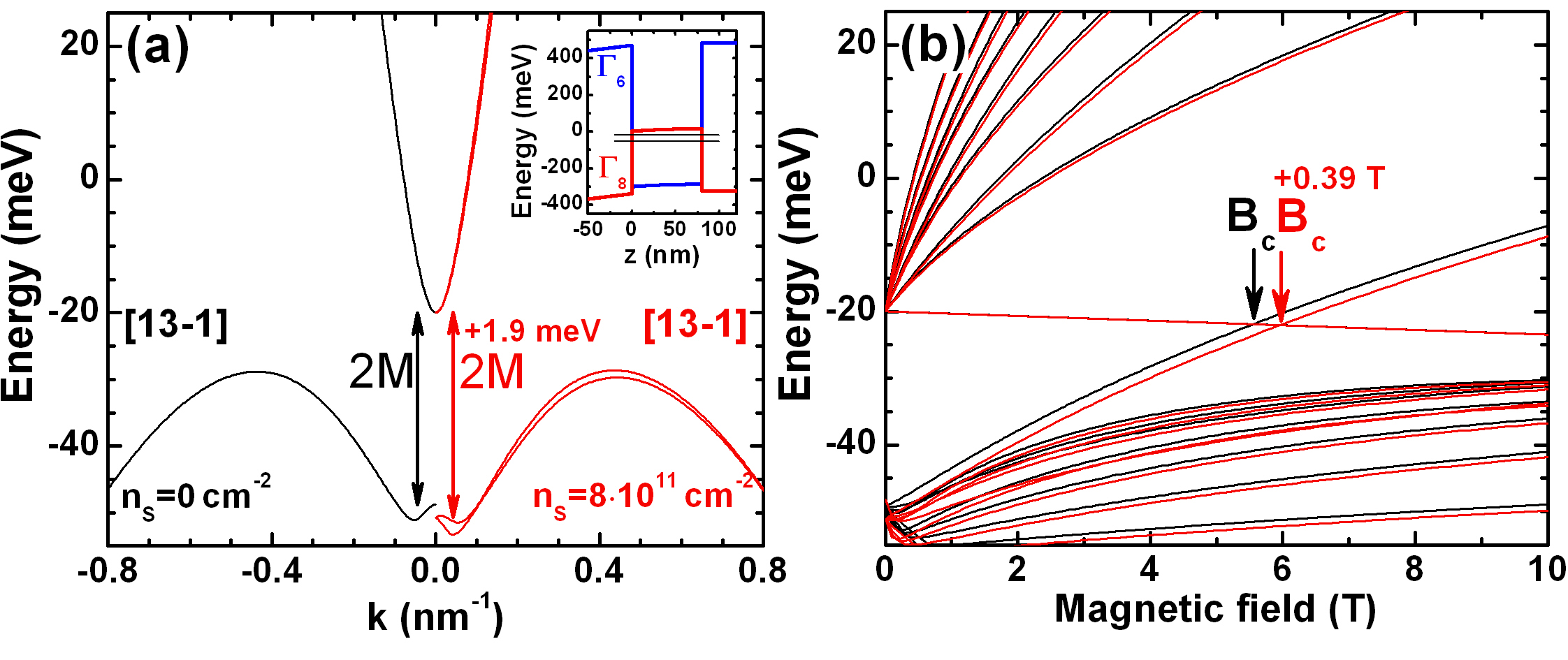} % Here is how to import EPS art
\caption{\label{Fig:SM6} Band structure (a) and Landau levels (b) of 8-nm-wide HgTe/Cd$_{0.7}$Hg$_{0.3}$Te QWs at $T=2$~K grown on (013) CdTe buffer. The black curves at both panels correspond to the rectangular QW profile ($n_S=0$), while the red curves represent self-consistent calculations made for $n_S=8\cdot10^{11}$ cm$^{-2}$ assuming that all the electrons come from one side of the QW. The self-consistent QW profile for the latter case is shown in the inset at the right panel.}
\end{figure}

Figure~\ref{Fig:SM6} compares the band structure and LLs for 8-nm-wide HgTe/Cd$_{0.7}$Hg$_{0.3}$Te QWs with rectangular ($n_S=0$) and asymmetrical profile ($n_S=8\cdot10^{11}$~cm$^{-2}$), assuming that all the electrons come from one side of the QW. The calculations were performed by neglecting the BIA and IIA effects. To calculate the LLs, we apply the axial approximation by omitting the warping terms in the Hamiltonian. As can be seen, the change in the energy difference between the \emph{E}1 and \emph{H}1 subbands at the $\Gamma$ point and the critical magnetic $B_c$ does not exceed 10\% for a change in concentration of $n_S=8\cdot10^{11}$~cm$^{-2}$.

We first note that in our experiment, the change in $n_S$ is less than $2\cdot10^{11}$~cm$^{-2}$ for all samples. Second, the calculations shown in Fig.~\ref{Fig:SM6} were made for pure asymmetric QWs, while the real samples have an almost symmetrical QW profile before illumination, as the sample barriers are symmetrically doped. Thus, the modification of the QW profile during illumination is indeed very small in our samples, and it cannot explain the experimental evolution of the parameters with concentration.

\begin{figure}
\includegraphics [width=1.0\columnwidth, keepaspectratio] {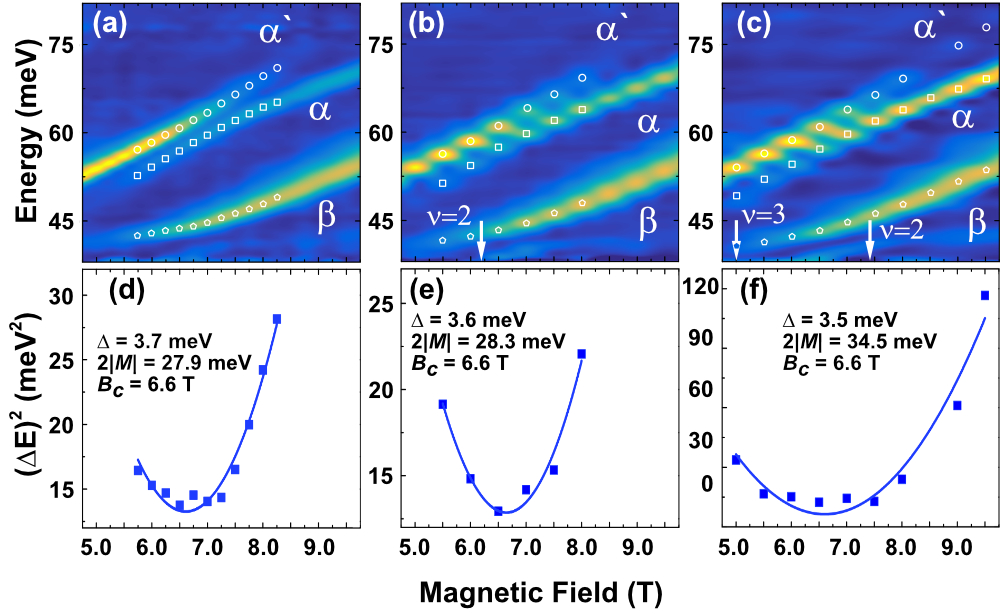} % Here is how to import EPS art
\caption{\label{Fig:SM1} False color maps showing $\alpha$ and $\alpha'$ inter-LL transitions in the 8~nm HgTe/Cd$_{0.41}$Hg$_{0.59}$Te QW (sample 101221) as a function of magnetic field, measured at different electron concentration $n_S$: (a)~$1.6\cdot10^{11}$~cm$^{-2}$, (b)~$3.0\cdot10^{11}$~cm$^{-2}$, (c)~$3.3\cdot10^{11}$~cm$^{-2}$. The symbols represent position of the magnetoabsorption lines, whose energies are used in the evaluation of $\Delta E$. (d)-(f) Square of the energy difference for $\alpha$ and $\alpha'$ transitions at the same concentrations as in the respective top panels. The solid curves are the fitting to Eq.~(\ref{eq:SM14}). The arrows indicate magnetic fields corresponding to the integer filling factor $\nu$.}
\includegraphics [width=1.0\columnwidth, keepaspectratio] {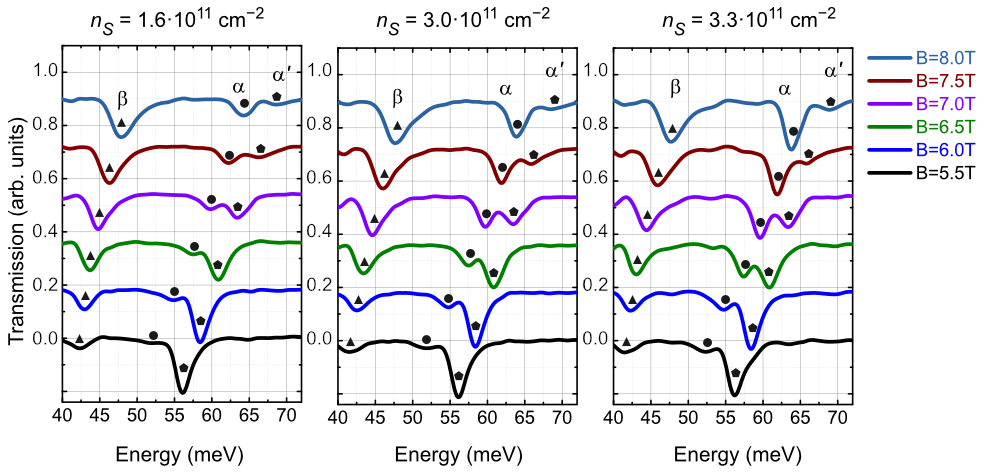} % Here is how to import EPS art
\caption{\label{Fig:SM4} Transmission spectra of sample 101221 in the same energy range as in Fig.~\ref{Fig:SM1} provided above. The resonant energies for $\alpha$, $\alpha'$ and $\beta$ transitions are marked by the symbols.}
\end{figure}

\subsection{D. Magnetotransmission spectra for all the samples}
Figure~\ref{Fig:SM1} provides magnetoabsorption spectra of the sample 101221 measured at different electron concentration $n_S$ varied by illumination of blue light emitting diode (see the main text). It is seen that the $\alpha$ and $\alpha'$ transitions are both present in the given magnetic field range as the filling factor of LLs $\nu<3$ for all $n_S$ values. On the contrary, for their observation, the $\beta$ and $\beta'$ transitions both require the stronger fields corresponding to $\nu<2$. Although the latter condition is fulfilled at some concentration value, significant distortion of the magnetoabsorption spectra in the vicinity of the reststrahlen band of GaAs substrate~\cite{SM8,SM9} makes clear observation of the $\beta'$ transition impossible in the sample 101221.

\begin{figure}
\includegraphics [width=0.98\columnwidth, keepaspectratio] {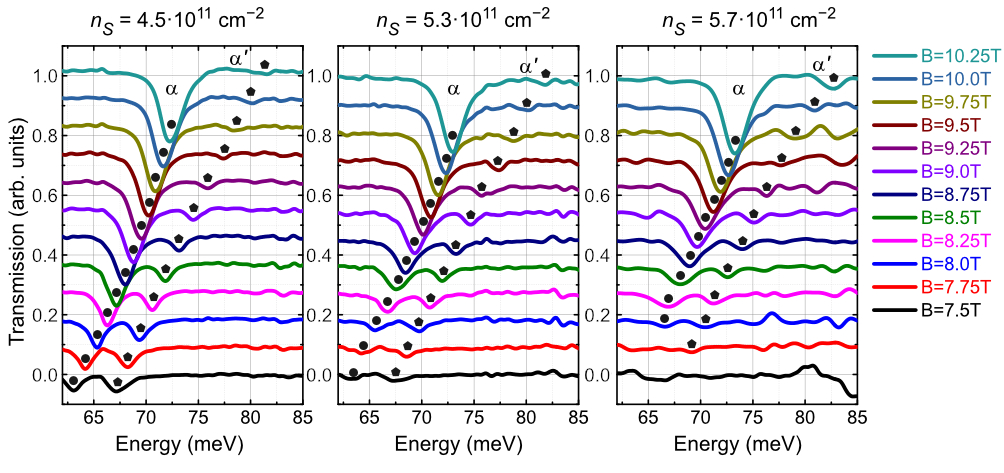} % Here is how to import EPS art
\caption{\label{Fig:SM2} Transmission spectra of sample 101109 in the same energy range as in Fig.~2 in the main text. The resonant energies for $\alpha$ and $\alpha'$ transitions are marked by the symbols.}
\includegraphics [width=0.98\columnwidth, keepaspectratio] {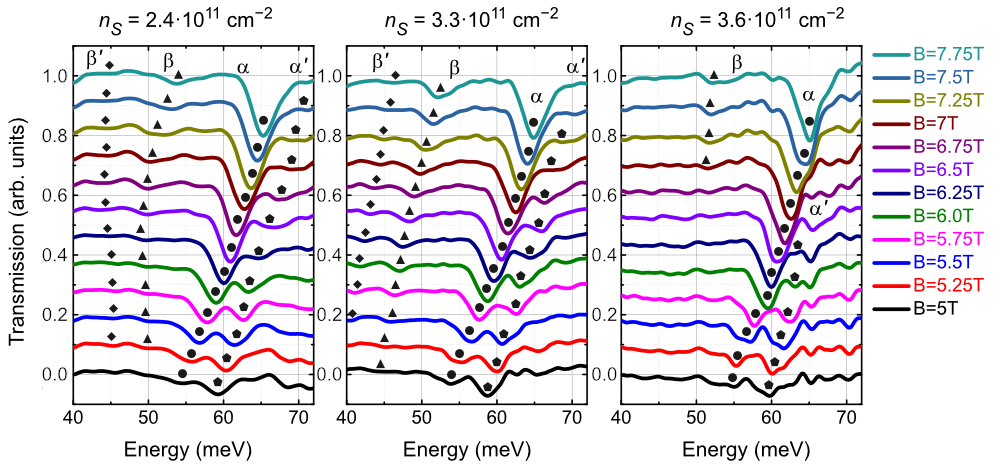} % Here is how to import EPS art
\caption{\label{Fig:SM3} Transmission spectra of sample 091223 in the same energy range as in Fig.~3 in the main text. The resonant energies for $\alpha$, $\alpha'$, $\beta$ and $\beta'$ transitions are marked by the symbols.}
\end{figure}

\begin{figure}
\includegraphics [width=1.0\columnwidth, keepaspectratio] {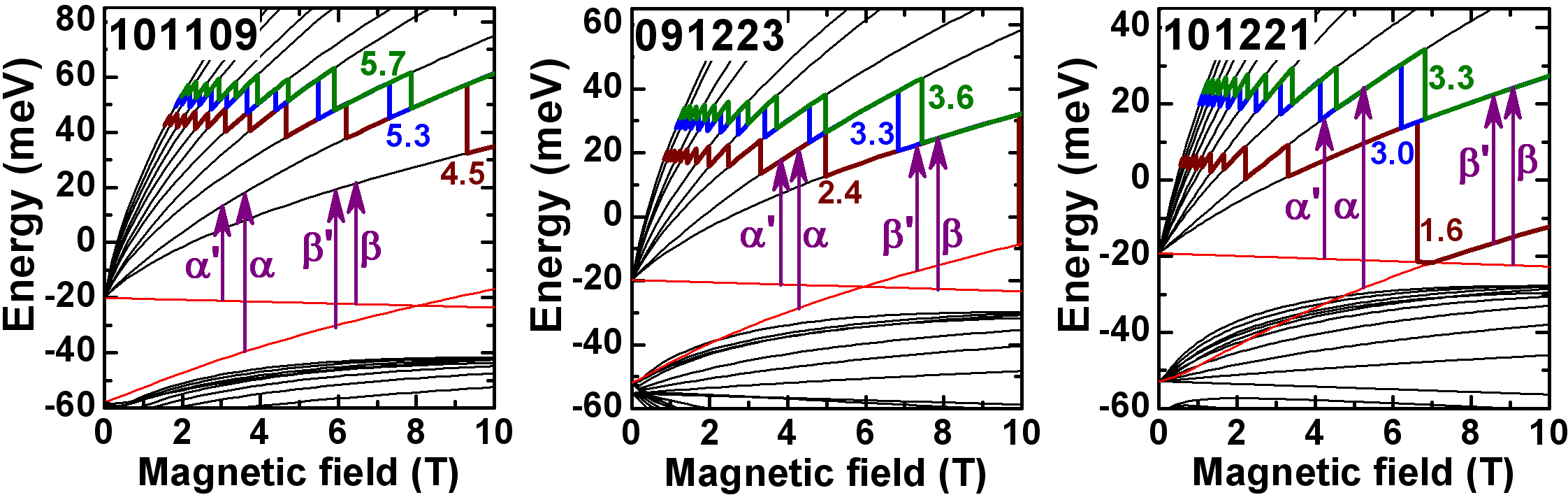} % Here is how to import EPS art
\caption{\label{Fig:SM5} Band structure Landau levels for three samples under study calculated within the 8-band k$\cdot$p Hamiltonian. The zero-mode LLs are shown in red. The arrows indicate $\alpha$, $\alpha'$, $\beta$ and $\beta'$ transitions. The colored bold curves represent the evolution of Fermi level with magnetic field at different concentration. The numbers show the values of $n_S$ in $10^{11}$~cm$^{-2}$.}
\end{figure}

Details of magnetotransmission spectra for all three samples under study are provided in Figs.~\ref{Fig:SM4}--\ref{Fig:SM3}. In order to illustrate the range of magnetic fields suitable for observation of $\alpha$, $\alpha'$, $\beta$ and $\beta'$ transitions in our samples, Fig.~\ref{Fig:SM5} shows the change of Fermi energy with magnetic fields at the experimental values of electron concentration $n_S$. The calculations of Landau levels were performed for the symmetrical QW profiles (i.e. without the SIA effects) by neglecting the BIA and IIA effects. As in the main text, we applied the axial approximation by omitting the warping terms in the Hamiltonian~\cite{SM10}.

%\bibliography{AnticrossingSM2}

\begin{thebibliography}{50}%
\makeatletter
\providecommand \@ifxundefined [1]{%
 \@ifx{#1\undefined}
}%
\providecommand \@ifnum [1]{%
 \ifnum #1\expandafter \@firstoftwo
 \else \expandafter \@secondoftwo
 \fi
}%
\providecommand \@ifx [1]{%
 \ifx #1\expandafter \@firstoftwo
 \else \expandafter \@secondoftwo
 \fi
}%
\providecommand \natexlab [1]{#1}%
\providecommand \enquote  [1]{``#1''}%
\providecommand \bibnamefont  [1]{#1}%
\providecommand \bibfnamefont [1]{#1}%
\providecommand \citenamefont [1]{#1}%
\providecommand \href@noop [0]{\@secondoftwo}%
\providecommand \href [0]{\begingroup \@sanitize@url \@href}%
\providecommand \@href[1]{\@@startlink{#1}\@@href}%
\providecommand \@@href[1]{\endgroup#1\@@endlink}%
\providecommand \@sanitize@url [0]{\catcode `\\12\catcode `\$12\catcode
  `\&12\catcode `\#12\catcode `\^12\catcode `\_12\catcode `\%12\relax}%
\providecommand \@@startlink[1]{}%
\providecommand \@@endlink[0]{}%
\providecommand \url  [0]{\begingroup\@sanitize@url \@url }%
\providecommand \@url [1]{\endgroup\@href {#1}{\urlprefix }}%
\providecommand \urlprefix  [0]{URL }%
\providecommand \Eprint [0]{\href }%
\providecommand \doibase [0]{http://dx.doi.org/}%
\providecommand \selectlanguage [0]{\@gobble}%
\providecommand \bibinfo  [0]{\@secondoftwo}%
\providecommand \bibfield  [0]{\@secondoftwo}%
\providecommand \translation [1]{[#1]}%
\providecommand \BibitemOpen [0]{}%
\providecommand \bibitemStop [0]{}%
\providecommand \bibitemNoStop [0]{.\EOS\space}%
\providecommand \EOS [0]{\spacefactor3000\relax}%
\providecommand \BibitemShut  [1]{\csname bibitem#1\endcsname}%
\let\auto@bib@innerbib\@empty
%</preamble>
\bibitem [{\citenamefont {Gerchikov}\ and\ \citenamefont
  {Subashiev}(1990)}]{A1}%
  \BibitemOpen
  \bibfield  {author} {\bibinfo {author} {\bibfnamefont {L.~G.}\ \bibnamefont
  {Gerchikov}}\ and\ \bibinfo {author} {\bibfnamefont {A.~V.}\ \bibnamefont
  {Subashiev}},\ }\href {\doibase 10.1002/pssb.2221600207} {\bibfield
  {journal} {\bibinfo  {journal} {Phys. Status Solidi B}\ }\textbf {\bibinfo
  {volume} {160}},\ \bibinfo {pages} {443} (\bibinfo {year}
  {1990})}\BibitemShut {NoStop}%
\bibitem [{\citenamefont {Bernevig}\ \emph {et~al.}(2006)\citenamefont
  {Bernevig}, \citenamefont {Hughes},\ and\ \citenamefont {Zhang}}]{A2}%
  \BibitemOpen
  \bibfield  {author} {\bibinfo {author} {\bibfnamefont {B.~A.}\ \bibnamefont
  {Bernevig}}, \bibinfo {author} {\bibfnamefont {T.~L.}\ \bibnamefont
  {Hughes}}, \ and\ \bibinfo {author} {\bibfnamefont {S.-C.}\ \bibnamefont
  {Zhang}},\ }\href {\doibase 10.1126/science.1133734} {\bibfield  {journal}
  {\bibinfo  {journal} {Science}\ }\textbf {\bibinfo {volume} {314}},\ \bibinfo
  {pages} {1757} (\bibinfo {year} {2006})}\BibitemShut {NoStop}%
\bibitem [{\citenamefont {K\"{o}nig}\ \emph {et~al.}(2007)\citenamefont
  {K\"{o}nig}, \citenamefont {Wiedmann}, \citenamefont {Br\"{u}ne},
  \citenamefont {Roth}, \citenamefont {Buhmann}, \citenamefont {Molenkamp},
  \citenamefont {Qi},\ and\ \citenamefont {Zhang}}]{A3}%
  \BibitemOpen
  \bibfield  {author} {\bibinfo {author} {\bibfnamefont {M.}~\bibnamefont
  {K\"{o}nig}}, \bibinfo {author} {\bibfnamefont {S.}~\bibnamefont {Wiedmann}},
  \bibinfo {author} {\bibfnamefont {C.}~\bibnamefont {Br\"{u}ne}}, \bibinfo
  {author} {\bibfnamefont {A.}~\bibnamefont {Roth}}, \bibinfo {author}
  {\bibfnamefont {H.}~\bibnamefont {Buhmann}}, \bibinfo {author} {\bibfnamefont
  {L.~W.}\ \bibnamefont {Molenkamp}}, \bibinfo {author} {\bibfnamefont {X.-L.}\
  \bibnamefont {Qi}}, \ and\ \bibinfo {author} {\bibfnamefont {S.-C.}\
  \bibnamefont {Zhang}},\ }\href {\doibase 10.1126/science.1148047} {\bibfield
  {journal} {\bibinfo  {journal} {Science}\ }\textbf {\bibinfo {volume}
  {318}},\ \bibinfo {pages} {766} (\bibinfo {year} {2007})}\BibitemShut
  {NoStop}%
\bibitem [{\citenamefont {B\"{u}ttner}\ \emph {et~al.}(2011)\citenamefont
  {B\"{u}ttner}, \citenamefont {Liu}, \citenamefont {Tkachov}, \citenamefont
  {Novik}, \citenamefont {Br\"{u}ne}, \citenamefont {Buhmann}, \citenamefont
  {Hankiewicz}, \citenamefont {Recher}, \citenamefont {Trauzettel},
  \citenamefont {Zhang},\ and\ \citenamefont {Molenkamp}}]{A4}%
  \BibitemOpen
  \bibfield  {author} {\bibinfo {author} {\bibfnamefont {B.}~\bibnamefont
  {B\"{u}ttner}}, \bibinfo {author} {\bibfnamefont {C.}~\bibnamefont {Liu}},
  \bibinfo {author} {\bibfnamefont {G.}~\bibnamefont {Tkachov}}, \bibinfo
  {author} {\bibfnamefont {E.}~\bibnamefont {Novik}}, \bibinfo {author}
  {\bibfnamefont {C.}~\bibnamefont {Br\"{u}ne}}, \bibinfo {author}
  {\bibfnamefont {H.}~\bibnamefont {Buhmann}}, \bibinfo {author} {\bibfnamefont
  {E.}~\bibnamefont {Hankiewicz}}, \bibinfo {author} {\bibfnamefont
  {P.}~\bibnamefont {Recher}}, \bibinfo {author} {\bibfnamefont
  {B.}~\bibnamefont {Trauzettel}}, \bibinfo {author} {\bibfnamefont
  {S.}~\bibnamefont {Zhang}}, \ and\ \bibinfo {author} {\bibfnamefont
  {L.}~\bibnamefont {Molenkamp}},\ }\href {\doibase 10.1038/nphys1914}
  {\bibfield  {journal} {\bibinfo  {journal} {Nat. Phys.}\ }\textbf {\bibinfo
  {volume} {7}},\ \bibinfo {pages} {418} (\bibinfo {year} {2011})}\BibitemShut
  {NoStop}%
\bibitem [{\citenamefont {Kadykov}\ \emph {et~al.}(2018)\citenamefont
  {Kadykov}, \citenamefont {Krishtopenko}, \citenamefont {Jouault},
  \citenamefont {Desrat}, \citenamefont {Knap}, \citenamefont {Ruffenach},
  \citenamefont {Consejo}, \citenamefont {Torres}, \citenamefont {Morozov},
  \citenamefont {Mikhailov}, \citenamefont {Dvoretskii},\ and\ \citenamefont
  {Teppe}}]{A5}%
  \BibitemOpen
  \bibfield  {author} {\bibinfo {author} {\bibfnamefont {A.~M.}\ \bibnamefont
  {Kadykov}}, \bibinfo {author} {\bibfnamefont {S.~S.}\ \bibnamefont
  {Krishtopenko}}, \bibinfo {author} {\bibfnamefont {B.}~\bibnamefont
  {Jouault}}, \bibinfo {author} {\bibfnamefont {W.}~\bibnamefont {Desrat}},
  \bibinfo {author} {\bibfnamefont {W.}~\bibnamefont {Knap}}, \bibinfo {author}
  {\bibfnamefont {S.}~\bibnamefont {Ruffenach}}, \bibinfo {author}
  {\bibfnamefont {C.}~\bibnamefont {Consejo}}, \bibinfo {author} {\bibfnamefont
  {J.}~\bibnamefont {Torres}}, \bibinfo {author} {\bibfnamefont {S.~V.}\
  \bibnamefont {Morozov}}, \bibinfo {author} {\bibfnamefont {N.~N.}\
  \bibnamefont {Mikhailov}}, \bibinfo {author} {\bibfnamefont {S.~A.}\
  \bibnamefont {Dvoretskii}}, \ and\ \bibinfo {author} {\bibfnamefont
  {F.}~\bibnamefont {Teppe}},\ }\href {\doibase 10.1103/PhysRevLett.120.086401}
  {\bibfield  {journal} {\bibinfo  {journal} {Phys. Rev. Lett.}\ }\textbf
  {\bibinfo {volume} {120}},\ \bibinfo {pages} {086401} (\bibinfo {year}
  {2018})}\BibitemShut {NoStop}%
\bibitem [{\citenamefont {Ludwig}\ \emph {et~al.}(2014)\citenamefont {Ludwig},
  \citenamefont {Vasilyev}, \citenamefont {Mikhailov}, \citenamefont
  {Poumirol}, \citenamefont {Jiang}, \citenamefont {Vafek},\ and\ \citenamefont
  {Smirnov}}]{A6}%
  \BibitemOpen
  \bibfield  {author} {\bibinfo {author} {\bibfnamefont {J.}~\bibnamefont
  {Ludwig}}, \bibinfo {author} {\bibfnamefont {Y.~B.}\ \bibnamefont
  {Vasilyev}}, \bibinfo {author} {\bibfnamefont {N.~N.}\ \bibnamefont
  {Mikhailov}}, \bibinfo {author} {\bibfnamefont {J.~M.}\ \bibnamefont
  {Poumirol}}, \bibinfo {author} {\bibfnamefont {Z.}~\bibnamefont {Jiang}},
  \bibinfo {author} {\bibfnamefont {O.}~\bibnamefont {Vafek}}, \ and\ \bibinfo
  {author} {\bibfnamefont {D.}~\bibnamefont {Smirnov}},\ }\href {\doibase
  10.1103/PhysRevB.89.241406} {\bibfield  {journal} {\bibinfo  {journal} {Phys.
  Rev. B}\ }\textbf {\bibinfo {volume} {89}},\ \bibinfo {pages} {241406}
  (\bibinfo {year} {2014})}\BibitemShut {NoStop}%
\bibitem [{\citenamefont {Marcinkiewicz}\ \emph {et~al.}(2017)\citenamefont
  {Marcinkiewicz}, \citenamefont {Ruffenach}, \citenamefont {Krishtopenko},
  \citenamefont {Kadykov}, \citenamefont {Consejo}, \citenamefont {But},
  \citenamefont {Desrat}, \citenamefont {Knap}, \citenamefont {Torres},
  \citenamefont {Ikonnikov}, \citenamefont {Spirin}, \citenamefont {Morozov},
  \citenamefont {Gavrilenko}, \citenamefont {Mikhailov}, \citenamefont
  {Dvoretskii},\ and\ \citenamefont {Teppe}}]{A7}%
  \BibitemOpen
  \bibfield  {author} {\bibinfo {author} {\bibfnamefont {M.}~\bibnamefont
  {Marcinkiewicz}}, \bibinfo {author} {\bibfnamefont {S.}~\bibnamefont
  {Ruffenach}}, \bibinfo {author} {\bibfnamefont {S.~S.}\ \bibnamefont
  {Krishtopenko}}, \bibinfo {author} {\bibfnamefont {A.~M.}\ \bibnamefont
  {Kadykov}}, \bibinfo {author} {\bibfnamefont {C.}~\bibnamefont {Consejo}},
  \bibinfo {author} {\bibfnamefont {D.~B.}\ \bibnamefont {But}}, \bibinfo
  {author} {\bibfnamefont {W.}~\bibnamefont {Desrat}}, \bibinfo {author}
  {\bibfnamefont {W.}~\bibnamefont {Knap}}, \bibinfo {author} {\bibfnamefont
  {J.}~\bibnamefont {Torres}}, \bibinfo {author} {\bibfnamefont {A.~V.}\
  \bibnamefont {Ikonnikov}}, \bibinfo {author} {\bibfnamefont {K.~E.}\
  \bibnamefont {Spirin}}, \bibinfo {author} {\bibfnamefont {S.~V.}\
  \bibnamefont {Morozov}}, \bibinfo {author} {\bibfnamefont {V.~I.}\
  \bibnamefont {Gavrilenko}}, \bibinfo {author} {\bibfnamefont {N.~N.}\
  \bibnamefont {Mikhailov}}, \bibinfo {author} {\bibfnamefont {S.~A.}\
  \bibnamefont {Dvoretskii}}, \ and\ \bibinfo {author} {\bibfnamefont
  {F.}~\bibnamefont {Teppe}},\ }\href {\doibase 10.1103/PhysRevB.96.035405}
  {\bibfield  {journal} {\bibinfo  {journal} {Phys. Rev. B}\ }\textbf {\bibinfo
  {volume} {96}},\ \bibinfo {pages} {035405} (\bibinfo {year}
  {2017})}\BibitemShut {NoStop}%
\bibitem [{\citenamefont {Krishtopenko}\ \emph
  {et~al.}(2016{\natexlab{a}})\citenamefont {Krishtopenko}, \citenamefont
  {Yahniuk}, \citenamefont {But}, \citenamefont {Gavrilenko}, \citenamefont
  {Knap},\ and\ \citenamefont {Teppe}}]{A8}%
  \BibitemOpen
  \bibfield  {author} {\bibinfo {author} {\bibfnamefont {S.~S.}\ \bibnamefont
  {Krishtopenko}}, \bibinfo {author} {\bibfnamefont {I.}~\bibnamefont
  {Yahniuk}}, \bibinfo {author} {\bibfnamefont {D.~B.}\ \bibnamefont {But}},
  \bibinfo {author} {\bibfnamefont {V.~I.}\ \bibnamefont {Gavrilenko}},
  \bibinfo {author} {\bibfnamefont {W.}~\bibnamefont {Knap}}, \ and\ \bibinfo
  {author} {\bibfnamefont {F.}~\bibnamefont {Teppe}},\ }\href {\doibase
  10.1103/PhysRevB.94.245402} {\bibfield  {journal} {\bibinfo  {journal} {Phys.
  Rev. B}\ }\textbf {\bibinfo {volume} {94}},\ \bibinfo {pages} {245402}
  (\bibinfo {year} {2016}{\natexlab{a}})}\BibitemShut {NoStop}%
\bibitem [{\citenamefont {Wiedmann}\ \emph {et~al.}(2015)\citenamefont
  {Wiedmann}, \citenamefont {Jost}, \citenamefont {Thienel}, \citenamefont
  {Br\"une}, \citenamefont {Leubner}, \citenamefont {Buhmann}, \citenamefont
  {Molenkamp}, \citenamefont {Maan},\ and\ \citenamefont {Zeitler}}]{A9}%
  \BibitemOpen
  \bibfield  {author} {\bibinfo {author} {\bibfnamefont {S.}~\bibnamefont
  {Wiedmann}}, \bibinfo {author} {\bibfnamefont {A.}~\bibnamefont {Jost}},
  \bibinfo {author} {\bibfnamefont {C.}~\bibnamefont {Thienel}}, \bibinfo
  {author} {\bibfnamefont {C.}~\bibnamefont {Br\"une}}, \bibinfo {author}
  {\bibfnamefont {P.}~\bibnamefont {Leubner}}, \bibinfo {author} {\bibfnamefont
  {H.}~\bibnamefont {Buhmann}}, \bibinfo {author} {\bibfnamefont {L.~W.}\
  \bibnamefont {Molenkamp}}, \bibinfo {author} {\bibfnamefont {J.~C.}\
  \bibnamefont {Maan}}, \ and\ \bibinfo {author} {\bibfnamefont
  {U.}~\bibnamefont {Zeitler}},\ }\href {\doibase 10.1103/PhysRevB.91.205311}
  {\bibfield  {journal} {\bibinfo  {journal} {Phys. Rev. B}\ }\textbf {\bibinfo
  {volume} {91}},\ \bibinfo {pages} {205311} (\bibinfo {year}
  {2015})}\BibitemShut {NoStop}%
\bibitem [{\citenamefont {Ikonnikov}\ \emph {et~al.}(2016)\citenamefont
  {Ikonnikov}, \citenamefont {Krishtopenko}, \citenamefont {Drachenko},
  \citenamefont {Goiran}, \citenamefont {Zholudev}, \citenamefont {Platonov},
  \citenamefont {Kudasov}, \citenamefont {Korshunov}, \citenamefont {Maslov},
  \citenamefont {Makarov}, \citenamefont {Surdin}, \citenamefont {Philippov},
  \citenamefont {Marcinkiewicz}, \citenamefont {Ruffenach}, \citenamefont
  {Teppe}, \citenamefont {Knap}, \citenamefont {Mikhailov}, \citenamefont
  {Dvoretsky},\ and\ \citenamefont {Gavrilenko}}]{A10}%
  \BibitemOpen
  \bibfield  {author} {\bibinfo {author} {\bibfnamefont {A.~V.}\ \bibnamefont
  {Ikonnikov}}, \bibinfo {author} {\bibfnamefont {S.~S.}\ \bibnamefont
  {Krishtopenko}}, \bibinfo {author} {\bibfnamefont {O.}~\bibnamefont
  {Drachenko}}, \bibinfo {author} {\bibfnamefont {M.}~\bibnamefont {Goiran}},
  \bibinfo {author} {\bibfnamefont {M.~S.}\ \bibnamefont {Zholudev}}, \bibinfo
  {author} {\bibfnamefont {V.~V.}\ \bibnamefont {Platonov}}, \bibinfo {author}
  {\bibfnamefont {Y.~B.}\ \bibnamefont {Kudasov}}, \bibinfo {author}
  {\bibfnamefont {A.~S.}\ \bibnamefont {Korshunov}}, \bibinfo {author}
  {\bibfnamefont {D.~A.}\ \bibnamefont {Maslov}}, \bibinfo {author}
  {\bibfnamefont {I.~V.}\ \bibnamefont {Makarov}}, \bibinfo {author}
  {\bibfnamefont {O.~M.}\ \bibnamefont {Surdin}}, \bibinfo {author}
  {\bibfnamefont {A.~V.}\ \bibnamefont {Philippov}}, \bibinfo {author}
  {\bibfnamefont {M.}~\bibnamefont {Marcinkiewicz}}, \bibinfo {author}
  {\bibfnamefont {S.}~\bibnamefont {Ruffenach}}, \bibinfo {author}
  {\bibfnamefont {F.}~\bibnamefont {Teppe}}, \bibinfo {author} {\bibfnamefont
  {W.}~\bibnamefont {Knap}}, \bibinfo {author} {\bibfnamefont {N.~N.}\
  \bibnamefont {Mikhailov}}, \bibinfo {author} {\bibfnamefont {S.~A.}\
  \bibnamefont {Dvoretsky}}, \ and\ \bibinfo {author} {\bibfnamefont {V.~I.}\
  \bibnamefont {Gavrilenko}},\ }\href {\doibase 10.1103/PhysRevB.94.155421}
  {\bibfield  {journal} {\bibinfo  {journal} {Phys. Rev. B}\ }\textbf {\bibinfo
  {volume} {94}},\ \bibinfo {pages} {155421} (\bibinfo {year}
  {2016})}\BibitemShut {NoStop}%
\bibitem [{\citenamefont {Leubner}\ \emph {et~al.}(2016)\citenamefont
  {Leubner}, \citenamefont {Lunczer}, \citenamefont {Br\"une}, \citenamefont
  {Buhmann},\ and\ \citenamefont {Molenkamp}}]{A11}%
  \BibitemOpen
  \bibfield  {author} {\bibinfo {author} {\bibfnamefont {P.}~\bibnamefont
  {Leubner}}, \bibinfo {author} {\bibfnamefont {L.}~\bibnamefont {Lunczer}},
  \bibinfo {author} {\bibfnamefont {C.}~\bibnamefont {Br\"une}}, \bibinfo
  {author} {\bibfnamefont {H.}~\bibnamefont {Buhmann}}, \ and\ \bibinfo
  {author} {\bibfnamefont {L.~W.}\ \bibnamefont {Molenkamp}},\ }\href {\doibase
  10.1103/PhysRevLett.117.086403} {\bibfield  {journal} {\bibinfo  {journal}
  {Phys. Rev. Lett.}\ }\textbf {\bibinfo {volume} {117}},\ \bibinfo {pages}
  {086403} (\bibinfo {year} {2016})}\BibitemShut {NoStop}%
\bibitem [{\citenamefont {Yahniuk}\ \emph {et~al.}(2019)\citenamefont
  {Yahniuk}, \citenamefont {Krishtopenko}, \citenamefont {Grabecki},
  \citenamefont {Jouault}, \citenamefont {Consejo}, \citenamefont {Desrat},
  \citenamefont {Majewicz}, \citenamefont {Kadykov}, \citenamefont {Spirin},
  \citenamefont {Gavrilenko}, \citenamefont {Mikhailov}, \citenamefont
  {Dvoretsky}, \citenamefont {But}, \citenamefont {Teppe}, \citenamefont
  {Wrobel}, \citenamefont {Cywinski}, \citenamefont {Kret}, \citenamefont
  {Dietl},\ and\ \citenamefont {Knap}}]{A12}%
  \BibitemOpen
  \bibfield  {author} {\bibinfo {author} {\bibfnamefont {I.}~\bibnamefont
  {Yahniuk}}, \bibinfo {author} {\bibfnamefont {S.~S.}\ \bibnamefont
  {Krishtopenko}}, \bibinfo {author} {\bibfnamefont {G.}~\bibnamefont
  {Grabecki}}, \bibinfo {author} {\bibfnamefont {B.}~\bibnamefont {Jouault}},
  \bibinfo {author} {\bibfnamefont {C.}~\bibnamefont {Consejo}}, \bibinfo
  {author} {\bibfnamefont {W.}~\bibnamefont {Desrat}}, \bibinfo {author}
  {\bibfnamefont {M.}~\bibnamefont {Majewicz}}, \bibinfo {author}
  {\bibfnamefont {A.~M.}\ \bibnamefont {Kadykov}}, \bibinfo {author}
  {\bibfnamefont {K.~E.}\ \bibnamefont {Spirin}}, \bibinfo {author}
  {\bibfnamefont {V.~I.}\ \bibnamefont {Gavrilenko}}, \bibinfo {author}
  {\bibfnamefont {N.~N.}\ \bibnamefont {Mikhailov}}, \bibinfo {author}
  {\bibfnamefont {S.~A.}\ \bibnamefont {Dvoretsky}}, \bibinfo {author}
  {\bibfnamefont {D.~B.}\ \bibnamefont {But}}, \bibinfo {author} {\bibfnamefont
  {F.}~\bibnamefont {Teppe}}, \bibinfo {author} {\bibfnamefont
  {J.}~\bibnamefont {Wrobel}}, \bibinfo {author} {\bibfnamefont
  {G.}~\bibnamefont {Cywinski}}, \bibinfo {author} {\bibfnamefont
  {S.}~\bibnamefont {Kret}}, \bibinfo {author} {\bibfnamefont {T.}~\bibnamefont
  {Dietl}}, \ and\ \bibinfo {author} {\bibfnamefont {W.}~\bibnamefont {Knap}},\
  }\href {\doibase 10.1038/s41535-019-0154-3} {\bibfield  {journal} {\bibinfo
  {journal} {npj Quantum Mater.}\ }\textbf {\bibinfo {volume} {4}},\ \bibinfo
  {pages} {13} (\bibinfo {year} {2019})}\BibitemShut {NoStop}%
\bibitem [{\citenamefont {Liu}\ \emph {et~al.}(2008)\citenamefont {Liu},
  \citenamefont {Hughes}, \citenamefont {Qi}, \citenamefont {Wang},\ and\
  \citenamefont {Zhang}}]{A13a}%
  \BibitemOpen
  \bibfield  {author} {\bibinfo {author} {\bibfnamefont {C.}~\bibnamefont
  {Liu}}, \bibinfo {author} {\bibfnamefont {T.~L.}\ \bibnamefont {Hughes}},
  \bibinfo {author} {\bibfnamefont {X.-L.}\ \bibnamefont {Qi}}, \bibinfo
  {author} {\bibfnamefont {K.}~\bibnamefont {Wang}}, \ and\ \bibinfo {author}
  {\bibfnamefont {S.-C.}\ \bibnamefont {Zhang}},\ }\href {\doibase
  10.1103/PhysRevLett.100.236601} {\bibfield  {journal} {\bibinfo  {journal}
  {Phys. Rev. Lett.}\ }\textbf {\bibinfo {volume} {100}},\ \bibinfo {pages}
  {236601} (\bibinfo {year} {2008})}\BibitemShut {NoStop}%
\bibitem [{\citenamefont {K\"{o}nig}\ \emph {et~al.}(2008)\citenamefont
  {K\"{o}nig}, \citenamefont {Buhmann}, \citenamefont {Molenkamp},
  \citenamefont {Hughes}, \citenamefont {Liu}, \citenamefont {Qi},\ and\
  \citenamefont {Zhang}}]{A13b}%
  \BibitemOpen
  \bibfield  {author} {\bibinfo {author} {\bibfnamefont {M.}~\bibnamefont
  {K\"{o}nig}}, \bibinfo {author} {\bibfnamefont {H.}~\bibnamefont {Buhmann}},
  \bibinfo {author} {\bibfnamefont {L.~W.}\ \bibnamefont {Molenkamp}}, \bibinfo
  {author} {\bibfnamefont {T.}~\bibnamefont {Hughes}}, \bibinfo {author}
  {\bibfnamefont {C.-X.}\ \bibnamefont {Liu}}, \bibinfo {author} {\bibfnamefont
  {X.-L.}\ \bibnamefont {Qi}}, \ and\ \bibinfo {author} {\bibfnamefont {S.-C.}\
  \bibnamefont {Zhang}},\ }\href {\doibase 10.1143/JPSJ.77.031007} {\bibfield
  {journal} {\bibinfo  {journal} {J. Phys. Soc. Jpn.}\ }\textbf {\bibinfo
  {volume} {77}},\ \bibinfo {pages} {031007} (\bibinfo {year}
  {2008})}\BibitemShut {NoStop}%
\bibitem [{\citenamefont {Durnev}\ and\ \citenamefont {Tarasenko}(2016)}]{A14}%
  \BibitemOpen
  \bibfield  {author} {\bibinfo {author} {\bibfnamefont {M.~V.}\ \bibnamefont
  {Durnev}}\ and\ \bibinfo {author} {\bibfnamefont {S.~A.}\ \bibnamefont
  {Tarasenko}},\ }\href {\doibase 10.1103/PhysRevB.93.075434} {\bibfield
  {journal} {\bibinfo  {journal} {Phys. Rev. B}\ }\textbf {\bibinfo {volume}
  {93}},\ \bibinfo {pages} {075434} (\bibinfo {year} {2016})}\BibitemShut
  {NoStop}%
\bibitem [{\citenamefont {Br\"{u}ne}\ \emph {et~al.}(2012)\citenamefont
  {Br\"{u}ne}, \citenamefont {Roth}, \citenamefont {Buhmann}, \citenamefont
  {Hankiewicz}, \citenamefont {Molenkamp}, \citenamefont {Maciejko},
  \citenamefont {Qi},\ and\ \citenamefont {Zhang}}]{A15}%
  \BibitemOpen
  \bibfield  {author} {\bibinfo {author} {\bibfnamefont {C.}~\bibnamefont
  {Br\"{u}ne}}, \bibinfo {author} {\bibfnamefont {A.}~\bibnamefont {Roth}},
  \bibinfo {author} {\bibfnamefont {H.}~\bibnamefont {Buhmann}}, \bibinfo
  {author} {\bibfnamefont {E.~M.}\ \bibnamefont {Hankiewicz}}, \bibinfo
  {author} {\bibfnamefont {L.~W.}\ \bibnamefont {Molenkamp}}, \bibinfo {author}
  {\bibfnamefont {J.}~\bibnamefont {Maciejko}}, \bibinfo {author}
  {\bibfnamefont {X.-L.}\ \bibnamefont {Qi}}, \ and\ \bibinfo {author}
  {\bibfnamefont {S.-C.}\ \bibnamefont {Zhang}},\ }\href {\doibase
  10.1038/nphys2322} {\bibfield  {journal} {\bibinfo  {journal} {Nat. Phys.}\
  }\textbf {\bibinfo {volume} {8}},\ \bibinfo {pages} {485} (\bibinfo {year}
  {2012})}\BibitemShut {NoStop}%
\bibitem [{\citenamefont {Olshanetsky}\ \emph {et~al.}(2018)\citenamefont
  {Olshanetsky}, \citenamefont {Kvon}, \citenamefont {Gusev}, \citenamefont
  {Mikhailov},\ and\ \citenamefont {Dvoretsky}}]{A16}%
  \BibitemOpen
  \bibfield  {author} {\bibinfo {author} {\bibfnamefont {E.}~\bibnamefont
  {Olshanetsky}}, \bibinfo {author} {\bibfnamefont {Z.}~\bibnamefont {Kvon}},
  \bibinfo {author} {\bibfnamefont {G.}~\bibnamefont {Gusev}}, \bibinfo
  {author} {\bibfnamefont {N.}~\bibnamefont {Mikhailov}}, \ and\ \bibinfo
  {author} {\bibfnamefont {S.}~\bibnamefont {Dvoretsky}},\ }\href {\doibase
  https://doi.org/10.1016/j.physe.2018.02.005} {\bibfield  {journal} {\bibinfo
  {journal} {Physica E Low. Dimens. Syst. Nanostruct.}\ }\textbf {\bibinfo
  {volume} {99}},\ \bibinfo {pages} {335 } (\bibinfo {year}
  {2018})}\BibitemShut {NoStop}%
\bibitem [{\citenamefont {Kadykov}\ \emph {et~al.}(2015)\citenamefont
  {Kadykov}, \citenamefont {Teppe}, \citenamefont {Consejo}, \citenamefont
  {Viti}, \citenamefont {Vitiello}, \citenamefont {Krishtopenko}, \citenamefont
  {Ruffenach}, \citenamefont {Morozov}, \citenamefont {Marcinkiewicz},
  \citenamefont {Desrat}, \citenamefont {Dyakonova}, \citenamefont {Knap},
  \citenamefont {Gavrilenko}, \citenamefont {Mikhailov},\ and\ \citenamefont
  {Dvoretsky}}]{A17}%
  \BibitemOpen
  \bibfield  {author} {\bibinfo {author} {\bibfnamefont {A.~M.}\ \bibnamefont
  {Kadykov}}, \bibinfo {author} {\bibfnamefont {F.}~\bibnamefont {Teppe}},
  \bibinfo {author} {\bibfnamefont {C.}~\bibnamefont {Consejo}}, \bibinfo
  {author} {\bibfnamefont {L.}~\bibnamefont {Viti}}, \bibinfo {author}
  {\bibfnamefont {M.~S.}\ \bibnamefont {Vitiello}}, \bibinfo {author}
  {\bibfnamefont {S.~S.}\ \bibnamefont {Krishtopenko}}, \bibinfo {author}
  {\bibfnamefont {S.}~\bibnamefont {Ruffenach}}, \bibinfo {author}
  {\bibfnamefont {S.~V.}\ \bibnamefont {Morozov}}, \bibinfo {author}
  {\bibfnamefont {M.}~\bibnamefont {Marcinkiewicz}}, \bibinfo {author}
  {\bibfnamefont {W.}~\bibnamefont {Desrat}}, \bibinfo {author} {\bibfnamefont
  {N.}~\bibnamefont {Dyakonova}}, \bibinfo {author} {\bibfnamefont
  {W.}~\bibnamefont {Knap}}, \bibinfo {author} {\bibfnamefont {V.~I.}\
  \bibnamefont {Gavrilenko}}, \bibinfo {author} {\bibfnamefont {N.~N.}\
  \bibnamefont {Mikhailov}}, \ and\ \bibinfo {author} {\bibfnamefont {S.~A.}\
  \bibnamefont {Dvoretsky}},\ }\href {\doibase 10.1063/1.4932943} {\bibfield
  {journal} {\bibinfo  {journal} {Appl. Phys. Lett.}\ }\textbf {\bibinfo
  {volume} {107}},\ \bibinfo {pages} {152101} (\bibinfo {year}
  {2015})}\BibitemShut {NoStop}%
\bibitem [{\citenamefont {Kadykov}\ \emph {et~al.}(2016)\citenamefont
  {Kadykov}, \citenamefont {Torres}, \citenamefont {Krishtopenko},
  \citenamefont {Consejo}, \citenamefont {Ruffenach}, \citenamefont
  {Marcinkiewicz}, \citenamefont {But}, \citenamefont {Knap}, \citenamefont
  {Morozov}, \citenamefont {Gavrilenko}, \citenamefont {Mikhailov},
  \citenamefont {Dvoretsky},\ and\ \citenamefont {Teppe}}]{A18}%
  \BibitemOpen
  \bibfield  {author} {\bibinfo {author} {\bibfnamefont {A.~M.}\ \bibnamefont
  {Kadykov}}, \bibinfo {author} {\bibfnamefont {J.}~\bibnamefont {Torres}},
  \bibinfo {author} {\bibfnamefont {S.~S.}\ \bibnamefont {Krishtopenko}},
  \bibinfo {author} {\bibfnamefont {C.}~\bibnamefont {Consejo}}, \bibinfo
  {author} {\bibfnamefont {S.}~\bibnamefont {Ruffenach}}, \bibinfo {author}
  {\bibfnamefont {M.}~\bibnamefont {Marcinkiewicz}}, \bibinfo {author}
  {\bibfnamefont {D.}~\bibnamefont {But}}, \bibinfo {author} {\bibfnamefont
  {W.}~\bibnamefont {Knap}}, \bibinfo {author} {\bibfnamefont {S.~V.}\
  \bibnamefont {Morozov}}, \bibinfo {author} {\bibfnamefont {V.~I.}\
  \bibnamefont {Gavrilenko}}, \bibinfo {author} {\bibfnamefont {N.~N.}\
  \bibnamefont {Mikhailov}}, \bibinfo {author} {\bibfnamefont {S.~A.}\
  \bibnamefont {Dvoretsky}}, \ and\ \bibinfo {author} {\bibfnamefont
  {F.}~\bibnamefont {Teppe}},\ }\href {\doibase 10.1063/1.4955018} {\bibfield
  {journal} {\bibinfo  {journal} {Appl. Phys. Lett.}\ }\textbf {\bibinfo
  {volume} {108}},\ \bibinfo {pages} {262102} (\bibinfo {year}
  {2016})}\BibitemShut {NoStop}%
\bibitem [{\citenamefont {Orlita}\ \emph {et~al.}(2011)\citenamefont {Orlita},
  \citenamefont {Masztalerz}, \citenamefont {Faugeras}, \citenamefont
  {Potemski}, \citenamefont {Novik}, \citenamefont {Br\"une}, \citenamefont
  {Buhmann},\ and\ \citenamefont {Molenkamp}}]{A19}%
  \BibitemOpen
  \bibfield  {author} {\bibinfo {author} {\bibfnamefont {M.}~\bibnamefont
  {Orlita}}, \bibinfo {author} {\bibfnamefont {K.}~\bibnamefont {Masztalerz}},
  \bibinfo {author} {\bibfnamefont {C.}~\bibnamefont {Faugeras}}, \bibinfo
  {author} {\bibfnamefont {M.}~\bibnamefont {Potemski}}, \bibinfo {author}
  {\bibfnamefont {E.~G.}\ \bibnamefont {Novik}}, \bibinfo {author}
  {\bibfnamefont {C.}~\bibnamefont {Br\"une}}, \bibinfo {author} {\bibfnamefont
  {H.}~\bibnamefont {Buhmann}}, \ and\ \bibinfo {author} {\bibfnamefont
  {L.~W.}\ \bibnamefont {Molenkamp}},\ }\href {\doibase
  10.1103/PhysRevB.83.115307} {\bibfield  {journal} {\bibinfo  {journal} {Phys.
  Rev. B}\ }\textbf {\bibinfo {volume} {83}},\ \bibinfo {pages} {115307}
  (\bibinfo {year} {2011})}\BibitemShut {NoStop}%
\bibitem [{\citenamefont {Zholudev}\ \emph {et~al.}(2012)\citenamefont
  {Zholudev}, \citenamefont {Teppe}, \citenamefont {Orlita}, \citenamefont
  {Consejo}, \citenamefont {Torres}, \citenamefont {Dyakonova}, \citenamefont
  {Czapkiewicz}, \citenamefont {Wr\'obel}, \citenamefont {Grabecki},
  \citenamefont {Mikhailov}, \citenamefont {Dvoretskii}, \citenamefont
  {Ikonnikov}, \citenamefont {Spirin}, \citenamefont {Aleshkin}, \citenamefont
  {Gavrilenko},\ and\ \citenamefont {Knap}}]{A20}%
  \BibitemOpen
  \bibfield  {author} {\bibinfo {author} {\bibfnamefont {M.}~\bibnamefont
  {Zholudev}}, \bibinfo {author} {\bibfnamefont {F.}~\bibnamefont {Teppe}},
  \bibinfo {author} {\bibfnamefont {M.}~\bibnamefont {Orlita}}, \bibinfo
  {author} {\bibfnamefont {C.}~\bibnamefont {Consejo}}, \bibinfo {author}
  {\bibfnamefont {J.}~\bibnamefont {Torres}}, \bibinfo {author} {\bibfnamefont
  {N.}~\bibnamefont {Dyakonova}}, \bibinfo {author} {\bibfnamefont
  {M.}~\bibnamefont {Czapkiewicz}}, \bibinfo {author} {\bibfnamefont
  {J.}~\bibnamefont {Wr\'obel}}, \bibinfo {author} {\bibfnamefont
  {G.}~\bibnamefont {Grabecki}}, \bibinfo {author} {\bibfnamefont
  {N.}~\bibnamefont {Mikhailov}}, \bibinfo {author} {\bibfnamefont
  {S.}~\bibnamefont {Dvoretskii}}, \bibinfo {author} {\bibfnamefont
  {A.}~\bibnamefont {Ikonnikov}}, \bibinfo {author} {\bibfnamefont
  {K.}~\bibnamefont {Spirin}}, \bibinfo {author} {\bibfnamefont
  {V.}~\bibnamefont {Aleshkin}}, \bibinfo {author} {\bibfnamefont
  {V.}~\bibnamefont {Gavrilenko}}, \ and\ \bibinfo {author} {\bibfnamefont
  {W.}~\bibnamefont {Knap}},\ }\href {\doibase 10.1103/PhysRevB.86.205420}
  {\bibfield  {journal} {\bibinfo  {journal} {Phys. Rev. B}\ }\textbf {\bibinfo
  {volume} {86}},\ \bibinfo {pages} {205420} (\bibinfo {year}
  {2012})}\BibitemShut {NoStop}%
\bibitem [{\citenamefont {Zholudev}\ \emph {et~al.}(2015)\citenamefont
  {Zholudev}, \citenamefont {Teppe}, \citenamefont {Morozov}, \citenamefont
  {Orlita}, \citenamefont {Consejo}, \citenamefont {Ruffenach}, \citenamefont
  {Knap}, \citenamefont {Gavrilenko}, \citenamefont {Dvoretskii},\ and\
  \citenamefont {Mikhailov}}]{A21}%
  \BibitemOpen
  \bibfield  {author} {\bibinfo {author} {\bibfnamefont {M.~S.}\ \bibnamefont
  {Zholudev}}, \bibinfo {author} {\bibfnamefont {F.}~\bibnamefont {Teppe}},
  \bibinfo {author} {\bibfnamefont {S.~V.}\ \bibnamefont {Morozov}}, \bibinfo
  {author} {\bibfnamefont {M.}~\bibnamefont {Orlita}}, \bibinfo {author}
  {\bibfnamefont {C.}~\bibnamefont {Consejo}}, \bibinfo {author} {\bibfnamefont
  {S.}~\bibnamefont {Ruffenach}}, \bibinfo {author} {\bibfnamefont
  {W.}~\bibnamefont {Knap}}, \bibinfo {author} {\bibfnamefont {V.~I.}\
  \bibnamefont {Gavrilenko}}, \bibinfo {author} {\bibfnamefont {S.~A.}\
  \bibnamefont {Dvoretskii}}, \ and\ \bibinfo {author} {\bibfnamefont {N.~N.}\
  \bibnamefont {Mikhailov}},\ }\href {\doibase 10.1134/S0021364014240175}
  {\bibfield  {journal} {\bibinfo  {journal} {JETP Lett.}\ }\textbf {\bibinfo
  {volume} {100}},\ \bibinfo {pages} {790} (\bibinfo {year}
  {2015})}\BibitemShut {NoStop}%
\bibitem [{\citenamefont {Bovkun}\ \emph {et~al.}(2019)\citenamefont {Bovkun},
  \citenamefont {Ikonnikov}, \citenamefont {Aleshkin}, \citenamefont {Spirin},
  \citenamefont {Gavrilenko}, \citenamefont {Mikhailov}, \citenamefont
  {Dvoretsky}, \citenamefont {Teppe}, \citenamefont {Piot}, \citenamefont
  {Potemski},\ and\ \citenamefont {Orlita}}]{A22}%
  \BibitemOpen
  \bibfield  {author} {\bibinfo {author} {\bibfnamefont {L.~S.}\ \bibnamefont
  {Bovkun}}, \bibinfo {author} {\bibfnamefont {A.~V.}\ \bibnamefont
  {Ikonnikov}}, \bibinfo {author} {\bibfnamefont {V.~Y.}\ \bibnamefont
  {Aleshkin}}, \bibinfo {author} {\bibfnamefont {K.~E.}\ \bibnamefont
  {Spirin}}, \bibinfo {author} {\bibfnamefont {V.~I.}\ \bibnamefont
  {Gavrilenko}}, \bibinfo {author} {\bibfnamefont {N.}~\bibnamefont
  {Mikhailov}}, \bibinfo {author} {\bibfnamefont {S.~A.}\ \bibnamefont
  {Dvoretsky}}, \bibinfo {author} {\bibfnamefont {F.}~\bibnamefont {Teppe}},
  \bibinfo {author} {\bibfnamefont {B.~A.}\ \bibnamefont {Piot}}, \bibinfo
  {author} {\bibfnamefont {M.}~\bibnamefont {Potemski}}, \ and\ \bibinfo
  {author} {\bibfnamefont {M.}~\bibnamefont {Orlita}},\ }\href {\doibase
  10.1088/1361-648x/aafdf0} {\bibfield  {journal} {\bibinfo  {journal} {J.
  Phys.: Condens. Matter}\ }\textbf {\bibinfo {volume} {31}},\ \bibinfo {pages}
  {145501} (\bibinfo {year} {2019})}\BibitemShut {NoStop}%
\bibitem [{\citenamefont {Tarasenko}\ \emph {et~al.}(2015)\citenamefont
  {Tarasenko}, \citenamefont {Durnev}, \citenamefont {Nestoklon}, \citenamefont
  {Ivchenko}, \citenamefont {Luo},\ and\ \citenamefont {Zunger}}]{A23}%
  \BibitemOpen
  \bibfield  {author} {\bibinfo {author} {\bibfnamefont {S.~A.}\ \bibnamefont
  {Tarasenko}}, \bibinfo {author} {\bibfnamefont {M.~V.}\ \bibnamefont
  {Durnev}}, \bibinfo {author} {\bibfnamefont {M.~O.}\ \bibnamefont
  {Nestoklon}}, \bibinfo {author} {\bibfnamefont {E.~L.}\ \bibnamefont
  {Ivchenko}}, \bibinfo {author} {\bibfnamefont {J.-W.}\ \bibnamefont {Luo}}, \
  and\ \bibinfo {author} {\bibfnamefont {A.}~\bibnamefont {Zunger}},\ }\href
  {\doibase 10.1103/PhysRevB.91.081302} {\bibfield  {journal} {\bibinfo
  {journal} {Phys. Rev. B}\ }\textbf {\bibinfo {volume} {91}},\ \bibinfo
  {pages} {081302} (\bibinfo {year} {2015})}\BibitemShut {NoStop}%
\bibitem [{\citenamefont {Minkov}\ \emph {et~al.}(2016)\citenamefont {Minkov},
  \citenamefont {Germanenko}, \citenamefont {Rut}, \citenamefont
  {Sherstobitov}, \citenamefont {Nestoklon}, \citenamefont {Dvoretski},\ and\
  \citenamefont {Mikhailov}}]{A24}%
  \BibitemOpen
  \bibfield  {author} {\bibinfo {author} {\bibfnamefont {G.~M.}\ \bibnamefont
  {Minkov}}, \bibinfo {author} {\bibfnamefont {A.~V.}\ \bibnamefont
  {Germanenko}}, \bibinfo {author} {\bibfnamefont {O.~E.}\ \bibnamefont {Rut}},
  \bibinfo {author} {\bibfnamefont {A.~A.}\ \bibnamefont {Sherstobitov}},
  \bibinfo {author} {\bibfnamefont {M.~O.}\ \bibnamefont {Nestoklon}}, \bibinfo
  {author} {\bibfnamefont {S.~A.}\ \bibnamefont {Dvoretski}}, \ and\ \bibinfo
  {author} {\bibfnamefont {N.~N.}\ \bibnamefont {Mikhailov}},\ }\href {\doibase
  10.1103/PhysRevB.93.155304} {\bibfield  {journal} {\bibinfo  {journal} {Phys.
  Rev. B}\ }\textbf {\bibinfo {volume} {93}},\ \bibinfo {pages} {155304}
  (\bibinfo {year} {2016})}\BibitemShut {NoStop}%
\bibitem [{\citenamefont {Bychkov}\ \emph {et~al.}(1981)\citenamefont
  {Bychkov}, \citenamefont {Iordanskii},\ and\ \citenamefont
  {Eliashberg}}]{A26}%
  \BibitemOpen
  \bibfield  {author} {\bibinfo {author} {\bibfnamefont {Y.~A.}\ \bibnamefont
  {Bychkov}}, \bibinfo {author} {\bibfnamefont {S.~V.}\ \bibnamefont
  {Iordanskii}}, \ and\ \bibinfo {author} {\bibfnamefont {G.~M.}\ \bibnamefont
  {Eliashberg}},\ }\href {http://jetpletters.ac.ru/ps/1502/article_22957.shtml}
  {\bibfield  {journal} {\bibinfo  {journal} {JETP Lett.}\ }\textbf {\bibinfo
  {volume} {33}},\ \bibinfo {pages} {143} (\bibinfo {year} {1981})}\BibitemShut
  {NoStop}%
\bibitem [{\citenamefont {Kallin}\ and\ \citenamefont {Halperin}(1984)}]{A27}%
  \BibitemOpen
  \bibfield  {author} {\bibinfo {author} {\bibfnamefont {C.}~\bibnamefont
  {Kallin}}\ and\ \bibinfo {author} {\bibfnamefont {B.~I.}\ \bibnamefont
  {Halperin}},\ }\href {\doibase 10.1103/PhysRevB.30.5655} {\bibfield
  {journal} {\bibinfo  {journal} {Phys. Rev. B}\ }\textbf {\bibinfo {volume}
  {30}},\ \bibinfo {pages} {5655} (\bibinfo {year} {1984})}\BibitemShut
  {NoStop}%
\bibitem [{\citenamefont {MacDonald}\ and\ \citenamefont {Kallin}(1989)}]{A29}%
  \BibitemOpen
  \bibfield  {author} {\bibinfo {author} {\bibfnamefont {A.~H.}\ \bibnamefont
  {MacDonald}}\ and\ \bibinfo {author} {\bibfnamefont {C.}~\bibnamefont
  {Kallin}},\ }\href {\doibase 10.1103/PhysRevB.40.5795} {\bibfield  {journal}
  {\bibinfo  {journal} {Phys. Rev. B}\ }\textbf {\bibinfo {volume} {40}},\
  \bibinfo {pages} {5795} (\bibinfo {year} {1989})}\BibitemShut {NoStop}%
\bibitem [{\citenamefont {Bychkov}\ and\ \citenamefont {Martinez}(2002)}]{A30}%
  \BibitemOpen
  \bibfield  {author} {\bibinfo {author} {\bibfnamefont {Y.~A.}\ \bibnamefont
  {Bychkov}}\ and\ \bibinfo {author} {\bibfnamefont {G.}~\bibnamefont
  {Martinez}},\ }\href {\doibase 10.1103/PhysRevB.66.193312} {\bibfield
  {journal} {\bibinfo  {journal} {Phys. Rev. B}\ }\textbf {\bibinfo {volume}
  {66}},\ \bibinfo {pages} {193312} (\bibinfo {year} {2002})}\BibitemShut
  {NoStop}%
\bibitem [{\citenamefont {Kohn}(1961)}]{A28}%
  \BibitemOpen
  \bibfield  {author} {\bibinfo {author} {\bibfnamefont {W.}~\bibnamefont
  {Kohn}},\ }\href {\doibase 10.1103/PhysRev.123.1242} {\bibfield  {journal}
  {\bibinfo  {journal} {Phys. Rev.}\ }\textbf {\bibinfo {volume} {123}},\
  \bibinfo {pages} {1242} (\bibinfo {year} {1961})}\BibitemShut {NoStop}%
\bibitem [{\citenamefont {Krishtopenko}(2013)}]{A33}%
  \BibitemOpen
  \bibfield  {author} {\bibinfo {author} {\bibfnamefont {S.~S.}\ \bibnamefont
  {Krishtopenko}},\ }\href {\doibase 10.1088/0953-8984/25/36/365602} {\bibfield
   {journal} {\bibinfo  {journal} {J. Phys.: Condens. Matter}\ }\textbf
  {\bibinfo {volume} {25}},\ \bibinfo {pages} {365602} (\bibinfo {year}
  {2013})}\BibitemShut {NoStop}%
\bibitem [{\citenamefont {Krishtopenko}\ \emph {et~al.}(2015)\citenamefont
  {Krishtopenko}, \citenamefont {Ikonnikov}, \citenamefont {Orlita},
  \citenamefont {Sadofyev}, \citenamefont {Goiran}, \citenamefont {Teppe},
  \citenamefont {Knap},\ and\ \citenamefont {Gavrilenko}}]{A34}%
  \BibitemOpen
  \bibfield  {author} {\bibinfo {author} {\bibfnamefont {S.~S.}\ \bibnamefont
  {Krishtopenko}}, \bibinfo {author} {\bibfnamefont {A.~V.}\ \bibnamefont
  {Ikonnikov}}, \bibinfo {author} {\bibfnamefont {M.}~\bibnamefont {Orlita}},
  \bibinfo {author} {\bibfnamefont {Y.~G.}\ \bibnamefont {Sadofyev}}, \bibinfo
  {author} {\bibfnamefont {M.}~\bibnamefont {Goiran}}, \bibinfo {author}
  {\bibfnamefont {F.}~\bibnamefont {Teppe}}, \bibinfo {author} {\bibfnamefont
  {W.}~\bibnamefont {Knap}}, \ and\ \bibinfo {author} {\bibfnamefont {V.~I.}\
  \bibnamefont {Gavrilenko}},\ }\href {\doibase 10.1063/1.4913927} {\bibfield
  {journal} {\bibinfo  {journal} {J. Appl. Phys.}\ }\textbf {\bibinfo {volume}
  {117}},\ \bibinfo {pages} {112813} (\bibinfo {year} {2015})}\BibitemShut
  {NoStop}%
\bibitem [{\citenamefont {Arimoto}\ \emph {et~al.}(2003)\citenamefont
  {Arimoto}, \citenamefont {Miura},\ and\ \citenamefont {Stradling}}]{A31}%
  \BibitemOpen
  \bibfield  {author} {\bibinfo {author} {\bibfnamefont {H.}~\bibnamefont
  {Arimoto}}, \bibinfo {author} {\bibfnamefont {N.}~\bibnamefont {Miura}}, \
  and\ \bibinfo {author} {\bibfnamefont {R.~A.}\ \bibnamefont {Stradling}},\
  }\href {\doibase 10.1103/PhysRevB.67.155319} {\bibfield  {journal} {\bibinfo
  {journal} {Phys. Rev. B}\ }\textbf {\bibinfo {volume} {67}},\ \bibinfo
  {pages} {155319} (\bibinfo {year} {2003})}\BibitemShut {NoStop}%
\bibitem [{\citenamefont {Hu}\ \emph {et~al.}(2003)\citenamefont {Hu},
  \citenamefont {Zehnder}, \citenamefont {Heyn},\ and\ \citenamefont
  {Heitmann}}]{A32}%
  \BibitemOpen
  \bibfield  {author} {\bibinfo {author} {\bibfnamefont {C.~M.}\ \bibnamefont
  {Hu}}, \bibinfo {author} {\bibfnamefont {C.}~\bibnamefont {Zehnder}},
  \bibinfo {author} {\bibfnamefont {C.}~\bibnamefont {Heyn}}, \ and\ \bibinfo
  {author} {\bibfnamefont {D.}~\bibnamefont {Heitmann}},\ }\href {\doibase
  10.1103/PhysRevB.67.201302} {\bibfield  {journal} {\bibinfo  {journal} {Phys.
  Rev. B}\ }\textbf {\bibinfo {volume} {67}},\ \bibinfo {pages} {201302}
  (\bibinfo {year} {2003})}\BibitemShut {NoStop}%
\bibitem [{\citenamefont {Krishtopenko}\ \emph {et~al.}()\citenamefont
  {Krishtopenko}, \citenamefont {Gavrilenko},\ and\ \citenamefont
  {Goiran}}]{A32b}%
  \BibitemOpen
  \bibfield  {author} {\bibinfo {author} {\bibfnamefont {S.~S.}\ \bibnamefont
  {Krishtopenko}}, \bibinfo {author} {\bibfnamefont {V.~I.}\ \bibnamefont
  {Gavrilenko}}, \ and\ \bibinfo {author} {\bibfnamefont {M.}~\bibnamefont
  {Goiran}},\ }\href {\doibase 10.4028/www.scientific.net/SSP.190.554}
  {\bibfield  {journal} {\bibinfo  {journal} {Solid State Phenomena}\ }\textbf
  {\bibinfo {volume} {190}},\ \bibinfo {pages} {554}}\BibitemShut {NoStop}%
\bibitem [{\citenamefont {Krishtopenko}\ and\ \citenamefont
  {Teppe}(2018)}]{A34b}%
  \BibitemOpen
  \bibfield  {author} {\bibinfo {author} {\bibfnamefont {S.~S.}\ \bibnamefont
  {Krishtopenko}}\ and\ \bibinfo {author} {\bibfnamefont {F.}~\bibnamefont
  {Teppe}},\ }\href {\doibase 10.1126/sciadv.aap7529} {\bibfield  {journal}
  {\bibinfo  {journal} {Sci. Adv.}\ }\textbf {\bibinfo {volume} {4}},\ \bibinfo
  {pages} {eaap7529} (\bibinfo {year} {2018})}\BibitemShut {NoStop}%
\bibitem [{\citenamefont {Jiang}\ \emph {et~al.}(2007)\citenamefont {Jiang},
  \citenamefont {Henriksen}, \citenamefont {Tung}, \citenamefont {Wang},
  \citenamefont {Schwartz}, \citenamefont {Han}, \citenamefont {Kim},\ and\
  \citenamefont {Stormer}}]{A35}%
  \BibitemOpen
  \bibfield  {author} {\bibinfo {author} {\bibfnamefont {Z.}~\bibnamefont
  {Jiang}}, \bibinfo {author} {\bibfnamefont {E.~A.}\ \bibnamefont
  {Henriksen}}, \bibinfo {author} {\bibfnamefont {L.~C.}\ \bibnamefont {Tung}},
  \bibinfo {author} {\bibfnamefont {Y.-J.}\ \bibnamefont {Wang}}, \bibinfo
  {author} {\bibfnamefont {M.~E.}\ \bibnamefont {Schwartz}}, \bibinfo {author}
  {\bibfnamefont {M.~Y.}\ \bibnamefont {Han}}, \bibinfo {author} {\bibfnamefont
  {P.}~\bibnamefont {Kim}}, \ and\ \bibinfo {author} {\bibfnamefont {H.~L.}\
  \bibnamefont {Stormer}},\ }\href {\doibase 10.1103/PhysRevLett.98.197403}
  {\bibfield  {journal} {\bibinfo  {journal} {Phys. Rev. Lett.}\ }\textbf
  {\bibinfo {volume} {98}},\ \bibinfo {pages} {197403} (\bibinfo {year}
  {2007})}\BibitemShut {NoStop}%
\bibitem [{\citenamefont {Henriksen}\ \emph {et~al.}(2010)\citenamefont
  {Henriksen}, \citenamefont {Cadden-Zimansky}, \citenamefont {Jiang},
  \citenamefont {Li}, \citenamefont {Tung}, \citenamefont {Schwartz},
  \citenamefont {Takita}, \citenamefont {Wang}, \citenamefont {Kim},\ and\
  \citenamefont {Stormer}}]{A36}%
  \BibitemOpen
  \bibfield  {author} {\bibinfo {author} {\bibfnamefont {E.~A.}\ \bibnamefont
  {Henriksen}}, \bibinfo {author} {\bibfnamefont {P.}~\bibnamefont
  {Cadden-Zimansky}}, \bibinfo {author} {\bibfnamefont {Z.}~\bibnamefont
  {Jiang}}, \bibinfo {author} {\bibfnamefont {Z.~Q.}\ \bibnamefont {Li}},
  \bibinfo {author} {\bibfnamefont {L.-C.}\ \bibnamefont {Tung}}, \bibinfo
  {author} {\bibfnamefont {M.~E.}\ \bibnamefont {Schwartz}}, \bibinfo {author}
  {\bibfnamefont {M.}~\bibnamefont {Takita}}, \bibinfo {author} {\bibfnamefont
  {Y.-J.}\ \bibnamefont {Wang}}, \bibinfo {author} {\bibfnamefont
  {P.}~\bibnamefont {Kim}}, \ and\ \bibinfo {author} {\bibfnamefont {H.~L.}\
  \bibnamefont {Stormer}},\ }\href {\doibase 10.1103/PhysRevLett.104.067404}
  {\bibfield  {journal} {\bibinfo  {journal} {Phys. Rev. Lett.}\ }\textbf
  {\bibinfo {volume} {104}},\ \bibinfo {pages} {067404} (\bibinfo {year}
  {2010})}\BibitemShut {NoStop}%
\bibitem [{\citenamefont {Faugeras}\ \emph {et~al.}(2015)\citenamefont
  {Faugeras}, \citenamefont {Berciaud}, \citenamefont {Leszczynski},
  \citenamefont {Henni}, \citenamefont {Nogajewski}, \citenamefont {Orlita},
  \citenamefont {Taniguchi}, \citenamefont {Watanabe}, \citenamefont
  {Forsythe}, \citenamefont {Kim}, \citenamefont {Jalil}, \citenamefont {Geim},
  \citenamefont {Basko},\ and\ \citenamefont {Potemski}}]{A37}%
  \BibitemOpen
  \bibfield  {author} {\bibinfo {author} {\bibfnamefont {C.}~\bibnamefont
  {Faugeras}}, \bibinfo {author} {\bibfnamefont {S.}~\bibnamefont {Berciaud}},
  \bibinfo {author} {\bibfnamefont {P.}~\bibnamefont {Leszczynski}}, \bibinfo
  {author} {\bibfnamefont {Y.}~\bibnamefont {Henni}}, \bibinfo {author}
  {\bibfnamefont {K.}~\bibnamefont {Nogajewski}}, \bibinfo {author}
  {\bibfnamefont {M.}~\bibnamefont {Orlita}}, \bibinfo {author} {\bibfnamefont
  {T.}~\bibnamefont {Taniguchi}}, \bibinfo {author} {\bibfnamefont
  {K.}~\bibnamefont {Watanabe}}, \bibinfo {author} {\bibfnamefont
  {C.}~\bibnamefont {Forsythe}}, \bibinfo {author} {\bibfnamefont
  {P.}~\bibnamefont {Kim}}, \bibinfo {author} {\bibfnamefont {R.}~\bibnamefont
  {Jalil}}, \bibinfo {author} {\bibfnamefont {A.~K.}\ \bibnamefont {Geim}},
  \bibinfo {author} {\bibfnamefont {D.~M.}\ \bibnamefont {Basko}}, \ and\
  \bibinfo {author} {\bibfnamefont {M.}~\bibnamefont {Potemski}},\ }\href
  {\doibase 10.1103/PhysRevLett.114.126804} {\bibfield  {journal} {\bibinfo
  {journal} {Phys. Rev. Lett.}\ }\textbf {\bibinfo {volume} {114}},\ \bibinfo
  {pages} {126804} (\bibinfo {year} {2015})}\BibitemShut {NoStop}%
\bibitem [{\citenamefont {Russell}\ \emph {et~al.}(2018)\citenamefont
  {Russell}, \citenamefont {Zhou}, \citenamefont {Taniguchi}, \citenamefont
  {Watanabe},\ and\ \citenamefont {Henriksen}}]{A38}%
  \BibitemOpen
  \bibfield  {author} {\bibinfo {author} {\bibfnamefont {B.~J.}\ \bibnamefont
  {Russell}}, \bibinfo {author} {\bibfnamefont {B.}~\bibnamefont {Zhou}},
  \bibinfo {author} {\bibfnamefont {T.}~\bibnamefont {Taniguchi}}, \bibinfo
  {author} {\bibfnamefont {K.}~\bibnamefont {Watanabe}}, \ and\ \bibinfo
  {author} {\bibfnamefont {E.~A.}\ \bibnamefont {Henriksen}},\ }\href {\doibase
  10.1103/PhysRevLett.120.047401} {\bibfield  {journal} {\bibinfo  {journal}
  {Phys. Rev. Lett.}\ }\textbf {\bibinfo {volume} {120}},\ \bibinfo {pages}
  {047401} (\bibinfo {year} {2018})}\BibitemShut {NoStop}%
\bibitem [{\citenamefont {Spirin}\ \emph {et~al.}(2012)\citenamefont {Spirin},
  \citenamefont {Kalinin}, \citenamefont {Krishtopenko}, \citenamefont
  {Maremyanin}, \citenamefont {Gavrilenko},\ and\ \citenamefont
  {Sadofyev}}]{A39}%
  \BibitemOpen
  \bibfield  {author} {\bibinfo {author} {\bibfnamefont {K.~E.}\ \bibnamefont
  {Spirin}}, \bibinfo {author} {\bibfnamefont {K.~P.}\ \bibnamefont {Kalinin}},
  \bibinfo {author} {\bibfnamefont {S.~S.}\ \bibnamefont {Krishtopenko}},
  \bibinfo {author} {\bibfnamefont {K.~V.}\ \bibnamefont {Maremyanin}},
  \bibinfo {author} {\bibfnamefont {V.~I.}\ \bibnamefont {Gavrilenko}}, \ and\
  \bibinfo {author} {\bibfnamefont {Y.~G.}\ \bibnamefont {Sadofyev}},\ }\href
  {\doibase 10.1134/S1063782612110206} {\bibfield  {journal} {\bibinfo
  {journal} {Semiconductors}\ }\textbf {\bibinfo {volume} {46}},\ \bibinfo
  {pages} {1396} (\bibinfo {year} {2012})}\BibitemShut {NoStop}%
\bibitem [{\citenamefont {Krishtopenko}\ \emph {et~al.}(2019)\citenamefont
  {Krishtopenko}, \citenamefont {Desrat}, \citenamefont {Spirin}, \citenamefont
  {Consejo}, \citenamefont {Ruffenach}, \citenamefont {Gonzalez-Posada},
  \citenamefont {Jouault}, \citenamefont {Knap}, \citenamefont {Maremyanin},
  \citenamefont {Gavrilenko}, \citenamefont {Boissier}, \citenamefont {Torres},
  \citenamefont {Zaknoune}, \citenamefont {Tourni\'e},\ and\ \citenamefont
  {Teppe}}]{A40}%
  \BibitemOpen
  \bibfield  {author} {\bibinfo {author} {\bibfnamefont {S.~S.}\ \bibnamefont
  {Krishtopenko}}, \bibinfo {author} {\bibfnamefont {W.}~\bibnamefont
  {Desrat}}, \bibinfo {author} {\bibfnamefont {K.~E.}\ \bibnamefont {Spirin}},
  \bibinfo {author} {\bibfnamefont {C.}~\bibnamefont {Consejo}}, \bibinfo
  {author} {\bibfnamefont {S.}~\bibnamefont {Ruffenach}}, \bibinfo {author}
  {\bibfnamefont {F.}~\bibnamefont {Gonzalez-Posada}}, \bibinfo {author}
  {\bibfnamefont {B.}~\bibnamefont {Jouault}}, \bibinfo {author} {\bibfnamefont
  {W.}~\bibnamefont {Knap}}, \bibinfo {author} {\bibfnamefont {K.~V.}\
  \bibnamefont {Maremyanin}}, \bibinfo {author} {\bibfnamefont {V.~I.}\
  \bibnamefont {Gavrilenko}}, \bibinfo {author} {\bibfnamefont
  {G.}~\bibnamefont {Boissier}}, \bibinfo {author} {\bibfnamefont
  {J.}~\bibnamefont {Torres}}, \bibinfo {author} {\bibfnamefont
  {M.}~\bibnamefont {Zaknoune}}, \bibinfo {author} {\bibfnamefont
  {E.}~\bibnamefont {Tourni\'e}}, \ and\ \bibinfo {author} {\bibfnamefont
  {F.}~\bibnamefont {Teppe}},\ }\href {\doibase 10.1103/PhysRevB.99.121405}
  {\bibfield  {journal} {\bibinfo  {journal} {Phys. Rev. B}\ }\textbf {\bibinfo
  {volume} {99}},\ \bibinfo {pages} {121405} (\bibinfo {year}
  {2019})}\BibitemShut {NoStop}%
\bibitem [{SM()}]{SM}%
  \BibitemOpen
  \href@noop {} {\bibinfo  {journal} {See Supplemental Materials, which also
  contain Refs.~[48-50], for a brief discussion of the joint effect of SIA, BIA
  and IIA on Landau levels within the Dirac-like BHZ model. Analysis of
  magnetoabsorption spectra of the sample 101221 and details of the spectra for
  the samples 101109 and 091223 are also provided therein}\ }\BibitemShut
  {NoStop}%
\bibitem [{\citenamefont {Schultz}\ \emph {et~al.}(1998)\citenamefont
  {Schultz}, \citenamefont {Merkt}, \citenamefont {Sonntag}, \citenamefont
  {R\"ossler}, \citenamefont {Winkler}, \citenamefont {Colin}, \citenamefont
  {Helgesen}, \citenamefont {Skauli},\ and\ \citenamefont {L\o{}vold}}]{A42}%
  \BibitemOpen
\bibfield  {journal} {  }\bibfield  {author} {\bibinfo {author} {\bibfnamefont
  {M.}~\bibnamefont {Schultz}}, \bibinfo {author} {\bibfnamefont
  {U.}~\bibnamefont {Merkt}}, \bibinfo {author} {\bibfnamefont
  {A.}~\bibnamefont {Sonntag}}, \bibinfo {author} {\bibfnamefont
  {U.}~\bibnamefont {R\"ossler}}, \bibinfo {author} {\bibfnamefont
  {R.}~\bibnamefont {Winkler}}, \bibinfo {author} {\bibfnamefont
  {T.}~\bibnamefont {Colin}}, \bibinfo {author} {\bibfnamefont
  {P.}~\bibnamefont {Helgesen}}, \bibinfo {author} {\bibfnamefont
  {T.}~\bibnamefont {Skauli}}, \ and\ \bibinfo {author} {\bibfnamefont
  {S.}~\bibnamefont {L\o{}vold}},\ }\href {\doibase 10.1103/PhysRevB.57.14772}
  {\bibfield  {journal} {\bibinfo  {journal} {Phys. Rev. B}\ }\textbf {\bibinfo
  {volume} {57}},\ \bibinfo {pages} {14772} (\bibinfo {year}
  {1998})}\BibitemShut {NoStop}%
\bibitem [{\citenamefont {Dvoretsky}\ \emph {et~al.}(2010)\citenamefont
  {Dvoretsky}, \citenamefont {Mikhailov}, \citenamefont {Sidorov},
  \citenamefont {Shvets}, \citenamefont {Danilov}, \citenamefont {Wittman},\
  and\ \citenamefont {Ganichev}}]{A41}%
  \BibitemOpen
  \bibfield  {author} {\bibinfo {author} {\bibfnamefont {S.}~\bibnamefont
  {Dvoretsky}}, \bibinfo {author} {\bibfnamefont {N.}~\bibnamefont
  {Mikhailov}}, \bibinfo {author} {\bibfnamefont {Y.}~\bibnamefont {Sidorov}},
  \bibinfo {author} {\bibfnamefont {V.}~\bibnamefont {Shvets}}, \bibinfo
  {author} {\bibfnamefont {S.}~\bibnamefont {Danilov}}, \bibinfo {author}
  {\bibfnamefont {B.}~\bibnamefont {Wittman}}, \ and\ \bibinfo {author}
  {\bibfnamefont {S.}~\bibnamefont {Ganichev}},\ }\href {\doibase
  10.1007/s11664-010-1191-7} {\bibfield  {journal} {\bibinfo  {journal} {J.
  Electron. Mater.}\ }\textbf {\bibinfo {volume} {39}},\ \bibinfo {pages} {918}
  (\bibinfo {year} {2010})}\BibitemShut {NoStop}%
\bibitem [{\citenamefont {Teppe}\ \emph {et~al.}(2016)\citenamefont {Teppe},
  \citenamefont {Marcinkiewicz}, \citenamefont {Krishtopenko}, \citenamefont
  {Ruffenach}, \citenamefont {Consejo}, \citenamefont {Kadykov}, \citenamefont
  {Desrat}, \citenamefont {But}, \citenamefont {Knap}, \citenamefont {Ludwig},
  \citenamefont {Moon}, \citenamefont {Smirnov}, \citenamefont {Orlita},
  \citenamefont {Jiang}, \citenamefont {Morozov}, \citenamefont {Gavrilenko},
  \citenamefont {Mikhailov},\ and\ \citenamefont {Dvoretskii}}]{A43}%
  \BibitemOpen
  \bibfield  {author} {\bibinfo {author} {\bibfnamefont {F.}~\bibnamefont
  {Teppe}}, \bibinfo {author} {\bibfnamefont {M.}~\bibnamefont
  {Marcinkiewicz}}, \bibinfo {author} {\bibfnamefont {S.~S.}\ \bibnamefont
  {Krishtopenko}}, \bibinfo {author} {\bibfnamefont {S.}~\bibnamefont
  {Ruffenach}}, \bibinfo {author} {\bibfnamefont {C.}~\bibnamefont {Consejo}},
  \bibinfo {author} {\bibfnamefont {A.~M.}\ \bibnamefont {Kadykov}}, \bibinfo
  {author} {\bibfnamefont {W.}~\bibnamefont {Desrat}}, \bibinfo {author}
  {\bibfnamefont {D.}~\bibnamefont {But}}, \bibinfo {author} {\bibfnamefont
  {W.}~\bibnamefont {Knap}}, \bibinfo {author} {\bibfnamefont {J.}~\bibnamefont
  {Ludwig}}, \bibinfo {author} {\bibfnamefont {S.}~\bibnamefont {Moon}},
  \bibinfo {author} {\bibfnamefont {D.}~\bibnamefont {Smirnov}}, \bibinfo
  {author} {\bibfnamefont {M.}~\bibnamefont {Orlita}}, \bibinfo {author}
  {\bibfnamefont {Z.}~\bibnamefont {Jiang}}, \bibinfo {author} {\bibfnamefont
  {S.~V.}\ \bibnamefont {Morozov}}, \bibinfo {author} {\bibfnamefont
  {V.}~\bibnamefont {Gavrilenko}}, \bibinfo {author} {\bibfnamefont {N.~N.}\
  \bibnamefont {Mikhailov}}, \ and\ \bibinfo {author} {\bibfnamefont {S.~A.}\
  \bibnamefont {Dvoretskii}},\ }\href {\doibase 10.1038/ncomms12576} {\bibfield
   {journal} {\bibinfo  {journal} {Nat. Commun.}\ }\textbf {\bibinfo {volume}
  {7}},\ \bibinfo {pages} {12576} (\bibinfo {year} {2016})}\BibitemShut
  {NoStop}%
\bibitem [{\citenamefont {Spirin}\ \emph {et~al.}(2019)\citenamefont {Spirin},
  \citenamefont {Gaponova}, \citenamefont {Gavrilenko}, \citenamefont
  {Mikhailov},\ and\ \citenamefont {Dvoretsky}}]{A40b}%
  \BibitemOpen
  \bibfield  {author} {\bibinfo {author} {\bibfnamefont {K.~E.}\ \bibnamefont
  {Spirin}}, \bibinfo {author} {\bibfnamefont {D.~M.}\ \bibnamefont
  {Gaponova}}, \bibinfo {author} {\bibfnamefont {V.~I.}\ \bibnamefont
  {Gavrilenko}}, \bibinfo {author} {\bibfnamefont {N.~N.}\ \bibnamefont
  {Mikhailov}}, \ and\ \bibinfo {author} {\bibfnamefont {S.~A.}\ \bibnamefont
  {Dvoretsky}},\ }\href {\doibase 10.1134/S106378261910021X} {\bibfield
  {journal} {\bibinfo  {journal} {Semiconductors}\ }\textbf {\bibinfo {volume}
  {53}},\ \bibinfo {pages} {1363} (\bibinfo {year} {2019})}\BibitemShut
  {NoStop}%
\bibitem [{\citenamefont {Krishtopenko}(2014)}]{SM6}%
  \BibitemOpen
  \bibfield  {author} {\bibinfo {author} {\bibfnamefont {S.~S.}\ \bibnamefont
  {Krishtopenko}},\ }\href {\doibase 10.1088/0268-1242/29/8/085005} {\bibfield
  {journal} {\bibinfo  {journal} {Semicond. Sci. Technol.}\ }\textbf {\bibinfo
  {volume} {29}},\ \bibinfo {pages} {085005} (\bibinfo {year}
  {2014})}\BibitemShut {NoStop}%
\bibitem [{\citenamefont {Krishtopenko}\ \emph
  {et~al.}(2016{\natexlab{b}})\citenamefont {Krishtopenko}, \citenamefont
  {Knap},\ and\ \citenamefont {Teppe}}]{SM7}%
  \BibitemOpen
  \bibfield  {author} {\bibinfo {author} {\bibfnamefont {S.~S.}\ \bibnamefont
  {Krishtopenko}}, \bibinfo {author} {\bibfnamefont {W.}~\bibnamefont {Knap}},
  \ and\ \bibinfo {author} {\bibfnamefont {F.}~\bibnamefont {Teppe}},\ }\href
  {\doibase 10.1038/srep30755} {\bibfield  {journal} {\bibinfo  {journal} {Sci.
  Rep.}\ }\textbf {\bibinfo {volume} {6}},\ \bibinfo {pages} {30755} (\bibinfo
  {year} {2016}{\natexlab{b}})}\BibitemShut {NoStop}%
\bibitem [{\citenamefont {Dobretsova}\ \emph {et~al.}(2019)\citenamefont
  {Dobretsova}, \citenamefont {Kvon}, \citenamefont {Krishtopenko},
  \citenamefont {Mikhailov},\ and\ \citenamefont {Dvoretsky}}]{SM14}%
  \BibitemOpen
  \bibfield  {author} {\bibinfo {author} {\bibfnamefont {A.~A.}\ \bibnamefont
  {Dobretsova}}, \bibinfo {author} {\bibfnamefont {Z.~D.}\ \bibnamefont
  {Kvon}}, \bibinfo {author} {\bibfnamefont {S.~S.}\ \bibnamefont
  {Krishtopenko}}, \bibinfo {author} {\bibfnamefont {N.~N.}\ \bibnamefont
  {Mikhailov}}, \ and\ \bibinfo {author} {\bibfnamefont {S.~A.}\ \bibnamefont
  {Dvoretsky}},\ }\href {\doibase 10.1063/1.5086405} {\bibfield  {journal}
  {\bibinfo  {journal} {J. Low Temp. Phys.}\ }\textbf {\bibinfo {volume}
  {45}},\ \bibinfo {pages} {159} (\bibinfo {year} {2019})}\BibitemShut
  {NoStop}%
\end{thebibliography}

\begin{thebibliography}{14}%
\makeatletter
\providecommand \@ifxundefined [1]{%
 \@ifx{#1\undefined}
}%
\providecommand \@ifnum [1]{%
 \ifnum #1\expandafter \@firstoftwo
 \else \expandafter \@secondoftwo
 \fi
}%
\providecommand \@ifx [1]{%
 \ifx #1\expandafter \@firstoftwo
 \else \expandafter \@secondoftwo
 \fi
}%
\providecommand \natexlab [1]{#1}%
\providecommand \enquote  [1]{``#1''}%
\providecommand \bibnamefont  [1]{#1}%
\providecommand \bibfnamefont [1]{#1}%
\providecommand \citenamefont [1]{#1}%
\providecommand \href@noop [0]{\@secondoftwo}%
\providecommand \href [0]{\begingroup \@sanitize@url \@href}%
\providecommand \@href[1]{\@@startlink{#1}\@@href}%
\providecommand \@@href[1]{\endgroup#1\@@endlink}%
\providecommand \@sanitize@url [0]{\catcode `\\12\catcode `\$12\catcode
  `\&12\catcode `\#12\catcode `\^12\catcode `\_12\catcode `\%12\relax}%
\providecommand \@@startlink[1]{}%
\providecommand \@@endlink[0]{}%
\providecommand \url  [0]{\begingroup\@sanitize@url \@url }%
\providecommand \@url [1]{\endgroup\@href {#1}{\urlprefix }}%
\providecommand \urlprefix  [0]{URL }%
\providecommand \Eprint [0]{\href }%
\providecommand \doibase [0]{http://dx.doi.org/}%
\providecommand \selectlanguage [0]{\@gobble}%
\providecommand \bibinfo  [0]{\@secondoftwo}%
\providecommand \bibfield  [0]{\@secondoftwo}%
\providecommand \translation [1]{[#1]}%
\providecommand \BibitemOpen [0]{}%
\providecommand \bibitemStop [0]{}%
\providecommand \bibitemNoStop [0]{.\EOS\space}%
\providecommand \EOS [0]{\spacefactor3000\relax}%
\providecommand \BibitemShut  [1]{\csname bibitem#1\endcsname}%
\let\auto@bib@innerbib\@empty
%</preamble>
\bibitem [{\citenamefont {Bernevig}\ \emph {et~al.}(2006)\citenamefont
  {Bernevig}, \citenamefont {Hughes},\ and\ \citenamefont {Zhang}}]{SM1}%
  \BibitemOpen
  \bibfield  {author} {\bibinfo {author} {\bibfnamefont {B.~A.}\ \bibnamefont
  {Bernevig}}, \bibinfo {author} {\bibfnamefont {T.~L.}\ \bibnamefont
  {Hughes}}, \ and\ \bibinfo {author} {\bibfnamefont {S.-C.}\ \bibnamefont
  {Zhang}},\ }\href {\doibase 10.1126/science.1133734} {\bibfield  {journal}
  {\bibinfo  {journal} {Science}\ }\textbf {\bibinfo {volume} {314}},\ \bibinfo
  {pages} {1757} (\bibinfo {year} {2006})}\BibitemShut {NoStop}%
\bibitem [{\citenamefont {Liu}\ \emph {et~al.}(2008)\citenamefont {Liu},
  \citenamefont {Hughes}, \citenamefont {Qi}, \citenamefont {Wang},\ and\
  \citenamefont {Zhang}}]{SM2}%
  \BibitemOpen
  \bibfield  {author} {\bibinfo {author} {\bibfnamefont {C.}~\bibnamefont
  {Liu}}, \bibinfo {author} {\bibfnamefont {T.~L.}\ \bibnamefont {Hughes}},
  \bibinfo {author} {\bibfnamefont {X.-L.}\ \bibnamefont {Qi}}, \bibinfo
  {author} {\bibfnamefont {K.}~\bibnamefont {Wang}}, \ and\ \bibinfo {author}
  {\bibfnamefont {S.-C.}\ \bibnamefont {Zhang}},\ }\href {\doibase
  10.1103/PhysRevLett.100.236601} {\bibfield  {journal} {\bibinfo  {journal}
  {Phys. Rev. Lett.}\ }\textbf {\bibinfo {volume} {100}},\ \bibinfo {pages}
  {236601} (\bibinfo {year} {2008})}\BibitemShut {NoStop}%
\bibitem [{\citenamefont {K\"{o}nig}\ \emph {et~al.}(2008)\citenamefont
  {K\"{o}nig}, \citenamefont {Buhmann}, \citenamefont {Molenkamp},
  \citenamefont {Hughes}, \citenamefont {Liu}, \citenamefont {Qi},\ and\
  \citenamefont {Zhang}}]{SM3}%
  \BibitemOpen
  \bibfield  {author} {\bibinfo {author} {\bibfnamefont {M.}~\bibnamefont
  {K\"{o}nig}}, \bibinfo {author} {\bibfnamefont {H.}~\bibnamefont {Buhmann}},
  \bibinfo {author} {\bibfnamefont {L.~W.}\ \bibnamefont {Molenkamp}}, \bibinfo
  {author} {\bibfnamefont {T.}~\bibnamefont {Hughes}}, \bibinfo {author}
  {\bibfnamefont {C.-X.}\ \bibnamefont {Liu}}, \bibinfo {author} {\bibfnamefont
  {X.-L.}\ \bibnamefont {Qi}}, \ and\ \bibinfo {author} {\bibfnamefont {S.-C.}\
  \bibnamefont {Zhang}},\ }\href {\doibase 10.1143/JPSJ.77.031007} {\bibfield
  {journal} {\bibinfo  {journal} {J. Phys. Soc. Jpn.}\ }\textbf {\bibinfo
  {volume} {77}},\ \bibinfo {pages} {031007} (\bibinfo {year}
  {2008})}\BibitemShut {NoStop}%
\bibitem [{\citenamefont {Tarasenko}\ \emph {et~al.}(2015)\citenamefont
  {Tarasenko}, \citenamefont {Durnev}, \citenamefont {Nestoklon}, \citenamefont
  {Ivchenko}, \citenamefont {Luo},\ and\ \citenamefont {Zunger}}]{SM4}%
  \BibitemOpen
  \bibfield  {author} {\bibinfo {author} {\bibfnamefont {S.~A.}\ \bibnamefont
  {Tarasenko}}, \bibinfo {author} {\bibfnamefont {M.~V.}\ \bibnamefont
  {Durnev}}, \bibinfo {author} {\bibfnamefont {M.~O.}\ \bibnamefont
  {Nestoklon}}, \bibinfo {author} {\bibfnamefont {E.~L.}\ \bibnamefont
  {Ivchenko}}, \bibinfo {author} {\bibfnamefont {J.-W.}\ \bibnamefont {Luo}}, \
  and\ \bibinfo {author} {\bibfnamefont {A.}~\bibnamefont {Zunger}},\ }\href
  {\doibase 10.1103/PhysRevB.91.081302} {\bibfield  {journal} {\bibinfo
  {journal} {Phys. Rev. B}\ }\textbf {\bibinfo {volume} {91}},\ \bibinfo
  {pages} {081302} (\bibinfo {year} {2015})}\BibitemShut {NoStop}%
\bibitem [{\citenamefont {K\"{o}nig}\ \emph {et~al.}(2007)\citenamefont
  {K\"{o}nig}, \citenamefont {Wiedmann}, \citenamefont {Br\"{u}ne},
  \citenamefont {Roth}, \citenamefont {Buhmann}, \citenamefont {Molenkamp},
  \citenamefont {Qi},\ and\ \citenamefont {Zhang}}]{SM5}%
  \BibitemOpen
  \bibfield  {author} {\bibinfo {author} {\bibfnamefont {M.}~\bibnamefont
  {K\"{o}nig}}, \bibinfo {author} {\bibfnamefont {S.}~\bibnamefont {Wiedmann}},
  \bibinfo {author} {\bibfnamefont {C.}~\bibnamefont {Br\"{u}ne}}, \bibinfo
  {author} {\bibfnamefont {A.}~\bibnamefont {Roth}}, \bibinfo {author}
  {\bibfnamefont {H.}~\bibnamefont {Buhmann}}, \bibinfo {author} {\bibfnamefont
  {L.~W.}\ \bibnamefont {Molenkamp}}, \bibinfo {author} {\bibfnamefont {X.-L.}\
  \bibnamefont {Qi}}, \ and\ \bibinfo {author} {\bibfnamefont {S.-C.}\
  \bibnamefont {Zhang}},\ }\href {\doibase 10.1126/science.1148047} {\bibfield
  {journal} {\bibinfo  {journal} {Science}\ }\textbf {\bibinfo {volume}
  {318}},\ \bibinfo {pages} {766} (\bibinfo {year} {2007})}\BibitemShut
  {NoStop}%
\bibitem [{\citenamefont {Krishtopenko}(2014)}]{SM6}%
  \BibitemOpen
  \bibfield  {author} {\bibinfo {author} {\bibfnamefont {S.~S.}\ \bibnamefont
  {Krishtopenko}},\ }\href {\doibase 10.1088/0268-1242/29/8/085005} {\bibfield
  {journal} {\bibinfo  {journal} {Semicond. Sci. Technol.}\ }\textbf {\bibinfo
  {volume} {29}},\ \bibinfo {pages} {085005} (\bibinfo {year}
  {2014})}\BibitemShut {NoStop}%
\bibitem [{\citenamefont {Krishtopenko}\ \emph
  {et~al.}(2016{\natexlab{a}})\citenamefont {Krishtopenko}, \citenamefont
  {Knap},\ and\ \citenamefont {Teppe}}]{SM7}%
  \BibitemOpen
  \bibfield  {author} {\bibinfo {author} {\bibfnamefont {S.~S.}\ \bibnamefont
  {Krishtopenko}}, \bibinfo {author} {\bibfnamefont {W.}~\bibnamefont {Knap}},
  \ and\ \bibinfo {author} {\bibfnamefont {F.}~\bibnamefont {Teppe}},\ }\href
  {\doibase 10.1038/srep30755} {\bibfield  {journal} {\bibinfo  {journal} {Sci.
  Rep.}\ }\textbf {\bibinfo {volume} {6}},\ \bibinfo {pages} {30755} (\bibinfo
  {year} {2016}{\natexlab{a}})}\BibitemShut {NoStop}%
\bibitem [{\citenamefont {Spirin}\ \emph {et~al.}(2012)\citenamefont {Spirin},
  \citenamefont {Kalinin}, \citenamefont {Krishtopenko}, \citenamefont
  {Maremyanin}, \citenamefont {Gavrilenko},\ and\ \citenamefont
  {Sadofyev}}]{SM11}%
  \BibitemOpen
  \bibfield  {author} {\bibinfo {author} {\bibfnamefont {K.~E.}\ \bibnamefont
  {Spirin}}, \bibinfo {author} {\bibfnamefont {K.~P.}\ \bibnamefont {Kalinin}},
  \bibinfo {author} {\bibfnamefont {S.~S.}\ \bibnamefont {Krishtopenko}},
  \bibinfo {author} {\bibfnamefont {K.~V.}\ \bibnamefont {Maremyanin}},
  \bibinfo {author} {\bibfnamefont {V.~I.}\ \bibnamefont {Gavrilenko}}, \ and\
  \bibinfo {author} {\bibfnamefont {Y.~G.}\ \bibnamefont {Sadofyev}},\ }\href
  {\doibase 10.1134/S1063782612110206} {\bibfield  {journal} {\bibinfo
  {journal} {Semiconductors}\ }\textbf {\bibinfo {volume} {46}},\ \bibinfo
  {pages} {1396} (\bibinfo {year} {2012})}\BibitemShut {NoStop}%
\bibitem [{\citenamefont {Krishtopenko}\ \emph {et~al.}(2019)\citenamefont
  {Krishtopenko}, \citenamefont {Desrat}, \citenamefont {Spirin}, \citenamefont
  {Consejo}, \citenamefont {Ruffenach}, \citenamefont {Gonzalez-Posada},
  \citenamefont {Jouault}, \citenamefont {Knap}, \citenamefont {Maremyanin},
  \citenamefont {Gavrilenko}, \citenamefont {Boissier}, \citenamefont {Torres},
  \citenamefont {Zaknoune}, \citenamefont {Tourni\'e},\ and\ \citenamefont
  {Teppe}}]{SM12}%
  \BibitemOpen
  \bibfield  {author} {\bibinfo {author} {\bibfnamefont {S.~S.}\ \bibnamefont
  {Krishtopenko}}, \bibinfo {author} {\bibfnamefont {W.}~\bibnamefont
  {Desrat}}, \bibinfo {author} {\bibfnamefont {K.~E.}\ \bibnamefont {Spirin}},
  \bibinfo {author} {\bibfnamefont {C.}~\bibnamefont {Consejo}}, \bibinfo
  {author} {\bibfnamefont {S.}~\bibnamefont {Ruffenach}}, \bibinfo {author}
  {\bibfnamefont {F.}~\bibnamefont {Gonzalez-Posada}}, \bibinfo {author}
  {\bibfnamefont {B.}~\bibnamefont {Jouault}}, \bibinfo {author} {\bibfnamefont
  {W.}~\bibnamefont {Knap}}, \bibinfo {author} {\bibfnamefont {K.~V.}\
  \bibnamefont {Maremyanin}}, \bibinfo {author} {\bibfnamefont {V.~I.}\
  \bibnamefont {Gavrilenko}}, \bibinfo {author} {\bibfnamefont
  {G.}~\bibnamefont {Boissier}}, \bibinfo {author} {\bibfnamefont
  {J.}~\bibnamefont {Torres}}, \bibinfo {author} {\bibfnamefont
  {M.}~\bibnamefont {Zaknoune}}, \bibinfo {author} {\bibfnamefont
  {E.}~\bibnamefont {Tourni\'e}}, \ and\ \bibinfo {author} {\bibfnamefont
  {F.}~\bibnamefont {Teppe}},\ }\href {\doibase 10.1103/PhysRevB.99.121405}
  {\bibfield  {journal} {\bibinfo  {journal} {Phys. Rev. B}\ }\textbf {\bibinfo
  {volume} {99}},\ \bibinfo {pages} {121405} (\bibinfo {year}
  {2019})}\BibitemShut {NoStop}%
\bibitem [{\citenamefont {Spirin}\ \emph {et~al.}(2019)\citenamefont {Spirin},
  \citenamefont {Gaponova}, \citenamefont {Gavrilenko}, \citenamefont
  {Mikhailov},\ and\ \citenamefont {Dvoretsky}}]{SM13}%
  \BibitemOpen
  \bibfield  {author} {\bibinfo {author} {\bibfnamefont {K.~E.}\ \bibnamefont
  {Spirin}}, \bibinfo {author} {\bibfnamefont {D.~M.}\ \bibnamefont
  {Gaponova}}, \bibinfo {author} {\bibfnamefont {V.~I.}\ \bibnamefont
  {Gavrilenko}}, \bibinfo {author} {\bibfnamefont {N.~N.}\ \bibnamefont
  {Mikhailov}}, \ and\ \bibinfo {author} {\bibfnamefont {S.~A.}\ \bibnamefont
  {Dvoretsky}},\ }\href {\doibase 10.1134/S106378261910021X} {\bibfield
  {journal} {\bibinfo  {journal} {Semiconductors}\ }\textbf {\bibinfo {volume}
  {53}},\ \bibinfo {pages} {1363} (\bibinfo {year} {2019})}\BibitemShut
  {NoStop}%
\bibitem [{\citenamefont {Dobretsova}\ \emph {et~al.}(2019)\citenamefont
  {Dobretsova}, \citenamefont {Kvon}, \citenamefont {Krishtopenko},
  \citenamefont {Mikhailov},\ and\ \citenamefont {Dvoretsky}}]{SM14}%
  \BibitemOpen
  \bibfield  {author} {\bibinfo {author} {\bibfnamefont {A.~A.}\ \bibnamefont
  {Dobretsova}}, \bibinfo {author} {\bibfnamefont {Z.~D.}\ \bibnamefont
  {Kvon}}, \bibinfo {author} {\bibfnamefont {S.~S.}\ \bibnamefont
  {Krishtopenko}}, \bibinfo {author} {\bibfnamefont {N.~N.}\ \bibnamefont
  {Mikhailov}}, \ and\ \bibinfo {author} {\bibfnamefont {S.~A.}\ \bibnamefont
  {Dvoretsky}},\ }\href {\doibase 10.1063/1.5086405} {\bibfield  {journal}
  {\bibinfo  {journal} {J. Low Temp. Phys.}\ }\textbf {\bibinfo {volume}
  {45}},\ \bibinfo {pages} {159} (\bibinfo {year} {2019})}\BibitemShut
  {NoStop}%
\bibitem [{\citenamefont {Orlita}\ \emph {et~al.}(2011)\citenamefont {Orlita},
  \citenamefont {Masztalerz}, \citenamefont {Faugeras}, \citenamefont
  {Potemski}, \citenamefont {Novik}, \citenamefont {Br\"une}, \citenamefont
  {Buhmann},\ and\ \citenamefont {Molenkamp}}]{SM8}%
  \BibitemOpen
  \bibfield  {author} {\bibinfo {author} {\bibfnamefont {M.}~\bibnamefont
  {Orlita}}, \bibinfo {author} {\bibfnamefont {K.}~\bibnamefont {Masztalerz}},
  \bibinfo {author} {\bibfnamefont {C.}~\bibnamefont {Faugeras}}, \bibinfo
  {author} {\bibfnamefont {M.}~\bibnamefont {Potemski}}, \bibinfo {author}
  {\bibfnamefont {E.~G.}\ \bibnamefont {Novik}}, \bibinfo {author}
  {\bibfnamefont {C.}~\bibnamefont {Br\"une}}, \bibinfo {author} {\bibfnamefont
  {H.}~\bibnamefont {Buhmann}}, \ and\ \bibinfo {author} {\bibfnamefont
  {L.~W.}\ \bibnamefont {Molenkamp}},\ }\href {\doibase
  10.1103/PhysRevB.83.115307} {\bibfield  {journal} {\bibinfo  {journal} {Phys.
  Rev. B}\ }\textbf {\bibinfo {volume} {83}},\ \bibinfo {pages} {115307}
  (\bibinfo {year} {2011})}\BibitemShut {NoStop}%
\bibitem [{\citenamefont {Zholudev}\ \emph {et~al.}(2012)\citenamefont
  {Zholudev}, \citenamefont {Teppe}, \citenamefont {Orlita}, \citenamefont
  {Consejo}, \citenamefont {Torres}, \citenamefont {Dyakonova}, \citenamefont
  {Czapkiewicz}, \citenamefont {Wr\'obel}, \citenamefont {Grabecki},
  \citenamefont {Mikhailov}, \citenamefont {Dvoretskii}, \citenamefont
  {Ikonnikov}, \citenamefont {Spirin}, \citenamefont {Aleshkin}, \citenamefont
  {Gavrilenko},\ and\ \citenamefont {Knap}}]{SM9}%
  \BibitemOpen
  \bibfield  {author} {\bibinfo {author} {\bibfnamefont {M.}~\bibnamefont
  {Zholudev}}, \bibinfo {author} {\bibfnamefont {F.}~\bibnamefont {Teppe}},
  \bibinfo {author} {\bibfnamefont {M.}~\bibnamefont {Orlita}}, \bibinfo
  {author} {\bibfnamefont {C.}~\bibnamefont {Consejo}}, \bibinfo {author}
  {\bibfnamefont {J.}~\bibnamefont {Torres}}, \bibinfo {author} {\bibfnamefont
  {N.}~\bibnamefont {Dyakonova}}, \bibinfo {author} {\bibfnamefont
  {M.}~\bibnamefont {Czapkiewicz}}, \bibinfo {author} {\bibfnamefont
  {J.}~\bibnamefont {Wr\'obel}}, \bibinfo {author} {\bibfnamefont
  {G.}~\bibnamefont {Grabecki}}, \bibinfo {author} {\bibfnamefont
  {N.}~\bibnamefont {Mikhailov}}, \bibinfo {author} {\bibfnamefont
  {S.}~\bibnamefont {Dvoretskii}}, \bibinfo {author} {\bibfnamefont
  {A.}~\bibnamefont {Ikonnikov}}, \bibinfo {author} {\bibfnamefont
  {K.}~\bibnamefont {Spirin}}, \bibinfo {author} {\bibfnamefont
  {V.}~\bibnamefont {Aleshkin}}, \bibinfo {author} {\bibfnamefont
  {V.}~\bibnamefont {Gavrilenko}}, \ and\ \bibinfo {author} {\bibfnamefont
  {W.}~\bibnamefont {Knap}},\ }\href {\doibase 10.1103/PhysRevB.86.205420}
  {\bibfield  {journal} {\bibinfo  {journal} {Phys. Rev. B}\ }\textbf {\bibinfo
  {volume} {86}},\ \bibinfo {pages} {205420} (\bibinfo {year}
  {2012})}\BibitemShut {NoStop}%
\bibitem [{\citenamefont {Krishtopenko}\ \emph
  {et~al.}(2016{\natexlab{b}})\citenamefont {Krishtopenko}, \citenamefont
  {Yahniuk}, \citenamefont {But}, \citenamefont {Gavrilenko}, \citenamefont
  {Knap},\ and\ \citenamefont {Teppe}}]{SM10}%
  \BibitemOpen
  \bibfield  {author} {\bibinfo {author} {\bibfnamefont {S.~S.}\ \bibnamefont
  {Krishtopenko}}, \bibinfo {author} {\bibfnamefont {I.}~\bibnamefont
  {Yahniuk}}, \bibinfo {author} {\bibfnamefont {D.~B.}\ \bibnamefont {But}},
  \bibinfo {author} {\bibfnamefont {V.~I.}\ \bibnamefont {Gavrilenko}},
  \bibinfo {author} {\bibfnamefont {W.}~\bibnamefont {Knap}}, \ and\ \bibinfo
  {author} {\bibfnamefont {F.}~\bibnamefont {Teppe}},\ }\href {\doibase
  10.1103/PhysRevB.94.245402} {\bibfield  {journal} {\bibinfo  {journal} {Phys.
  Rev. B}\ }\textbf {\bibinfo {volume} {94}},\ \bibinfo {pages} {245402}
  (\bibinfo {year} {2016}{\natexlab{b}})}\BibitemShut {NoStop}%
\end{thebibliography}

%

\end{document}